\documentclass[12pt, draftclsnofoot, onecolumn]{IEEEtran}
\usepackage{array}
\usepackage{color}
\usepackage{epsf}
\usepackage{times}
\usepackage{epsfig}
\usepackage{graphicx}
\usepackage{epstopdf}
\usepackage{hyperref}
\usepackage{amsmath}
\usepackage{amssymb}
\usepackage{amsxtra}
\usepackage{amsthm}
\usepackage{algorithmic}
\usepackage{algorithm}
\usepackage{caption}
\usepackage{subcaption}
\usepackage{authblk}
\usepackage{bbm}
\usepackage{comment}
\usepackage{soul}

\usepackage{xcolor,cite,etoolbox}
\makeatletter
\pretocmd\@bibitem{\color{black}\csname keycolor#1\endcsname}{}{\fail}
\newcommand\citecolor[1]{\@namedef{keycolor#1}{\color{blue}}}
\makeatother

\IEEEoverridecommandlockouts
\newtheorem{theorem}{Theorem}
\newtheorem{lemma}{Lemma}

\newtheorem{remark}{Remark}

\begin{document}
	\title{Coverage and Rate Analysis in Coexisting Terahertz and RF Finite Wireless Networks\\
	}
		\author{Nour~Kouzayha,~\IEEEmembership{Member,~IEEE,}
		Mustafa A.~Kishk,~\IEEEmembership{Member,~IEEE,}
		Hadi~Sarieddeen,~\IEEEmembership{Member,~IEEE,}
		Tareq Y. Al-Naffouri,~\IEEEmembership{Senior Member,~IEEE,}
		and~Mohamed-Slim Alouini,~\IEEEmembership{Fellow,~IEEE}
		\thanks{The authors are with the Division of Computer, Electrical and Mathematical Sciences and Engineering, King Abdullah University of Science and Technology, Thuwal, Saudi Arabia (e-mail: nour.kouzayha@kaust.edu.sa; mustafa.kishk@kaust.edu.sa; hadi.sarieddeen@kaust.edu.sa; tareq.alnaffouri@kaust.edu.sa; slim.alouini@kaust.edu.sa).}
	}
	\maketitle
	\vspace{-1.5cm}
\begin{abstract}
Wireless communications over Terahertz (THz)-band frequencies are vital enablers of ultra-high rate applications and services in sixth-generation ($6$G) networks. However, THz communications suffer from poor coverage because of inherent THz features such as high penetration losses, severe path loss, and significant molecular absorption. To surmount these critical challenges and fully exploit the THz band, we explore a coexisting radio frequency (RF) and THz finite indoor network in which THz small cells are deployed to provide high data rates, and RF macrocells are deployed to satisfy coverage requirements. Using stochastic geometry tools, we assess the performance of coexisting RF and THz networks in terms of coverage probability and average achievable rate. The accuracy of the analytical results is validated with Monte-Carlo simulations. Several insights are devised for accurate tuning and optimization of THz system parameters, including the fraction of THz access points (APs) to deploy, and the THz bias. The obtained results recognize a clear coverage/rate trade-off where a high fraction of THz AP improves the rate significantly but may degrade the coverage performance. Furthermore, the location of the user in the finite area highly affects the fraction of THz APs that optimizes the performance.
\end{abstract}

\section{Introduction}
Following the successful deployment of millimeter-wave (mmWave) technology in the fifth generation (5G) of wireless communications~\cite{Andrews}, Terahertz (THz) communications are being envisioned as critical enablers for alleviating spectrum scarcity and breaking the capacity limitation in the sixth generation (6G) of wireless networks~\cite{Sarieddeen,Elayan,rajatheva2020scoring}. Specifically, the ultra-wide THz band that ranges from $0.1$~THz to $10$~THz promises to support applications with high quality of service characteristics and terabits per seconds data rates. Furthermore, THz networks can realize highly secure communications and massive connectivity with plenty of available spectrum resources as more than $10$ billion devices are expected to be connected in the coming few years~\cite{Zhengquan}. The THz band also provides a remarkable potential for enabling accurate sensing and localization techniques~\cite{Sarieddeen,Lima9330512}.

Despite the vision and promise of the THz technology, the unique properties of THz wave propagation impose several challenges that hinder the efficient deployment of THz networks~\cite{Ian,sarieddeen2021overview}. For instance, compared to lower frequency bands, atmospheric effects can significantly degrade THz propagation and result in high spreading and molecular absorption losses \cite{tarboush2021teramimo}. Such losses decrease the THz transmission distance and the cell size, which requires accurate planning for THz network deployments~\cite{Jornet}. Moreover, high reflection and scattering losses are encountered at high frequencies, which causes the attenuation of non-line-of-sight (NLOS) rays as the power of the received signal becomes very low when reflected or scattered~\cite{Han,tarboush2021teramimo}. Furthermore, due to short wavelengths, THz communications are highly vulnerable to the existence of small blockages such as the user itself or moving humans in the environment~\cite{Akyildiz}. These blockages cause significant attenuation to THz propagation because of the high penetration loss, further decreasing the THz transmission range. To cope with the limited range of THz networks, ultra-dense deployments are considered. However, such deployments significantly increase the interference at users, limiting network density. The distinctive features of THz communications motivate the design and development of new solutions to address all these challenges and efficiently deploy THz networks~\cite{Chen,Faisal}.

Because of their limited bandwidth, sub-$6$~GHz technologies cannot cope with very high data-rate demands. Integrating small cells that operate in the THz band is fundamental to satisfying the increasing need for ultra-high data rates; keeping some sub-$6$~GHz cells can surpass the limited coverage of THz communications. This work presents a comprehensive analytical framework to derive the coverage probability and average achievable rates in a coexisting RF and THz finite indoor network. THz small cells are deployed to provide high data rates, and RF macrocells are deployed to satisfy coverage requirements.

\subsection{Related Work}
Stochastic geometry is used extensively for studying various aspects of wireless networks, characterizing their functionality, and understanding their operation~\cite{ElSawy}. In THz systems, although narrow THz beam widths result in less interference, excessive inter-cell interference is imposed because of the dense deployment of THz base stations to surpass the high path and molecular absorption losses. This further motivates the use of stochastic geometry thanks to its ability to introduce a mathematically compliant formulation for the inter-cell interference in the network analysis, which is incredibly challenging under other approaches. 

Using tools from stochastic geometry, several works are developed in the literature to study the performance of networks operating in the THz frequency band. To date, the majority of existing research works focus on studying the performance of THz-only networks~\cite{Kokkoniemi}. Furthermore, most of these works refer to the mean interference power and signal to interference and noise ratio ($\mathrm{SINR}$) and their moments, rather than characterizing the coverage probability and average rate~\cite{Petrov,Yao2,Wang,Dmitri}. In~\cite{Petrov,Yao2,Wang}, the performance of THz networks is studied using stochastic geometry while taking into consideration both the effect of indoor blockages and directional antennas. However, the reliability of these models is mediocre as they rely mainly on approximations. For instance, the authors in~\cite{Petrov} approximate the mean and variance of the interference and the $\mathrm{SINR}$ using the Taylor expansion. In~\cite{Yao2,Wang}, the mean interference is used instead of the instantaneous interference in deriving the coverage probability of the THz network. The work in~\cite{Humadi} uses the central limit theorem and the normal distribution to approximate the interference and obtain the coverage probability in a user-centric THz network where several base stations (BSs) cooperate to serve each user.  

Two-dimensional (2D) architectures are usually used in modeling sub-$6$~GHz networks to simplify the analysis and derive tractable performance metrics. While it may be acceptable to neglect the effect of vertical heights in sub-$6$~GHz networks because of the large transmission distance, such assumption is insufficient for dense THz networks with limited transmission ranges~\cite{Huq}. Recently, few papers in the literature have determined the coverage probability of a THz network in a three-dimensional (3D) indoor environment~\cite{Wu,Wu2,shafie,Shafie2} in which THz transmitters are mounted on the ceiling with fixed height to serve users. These works show the impact of different blockage types, including walls and moving humans, on the reliability performance of THz networks~\cite{shafie,Wu2}. While small scale fading is ignored in~\cite{shafie,Shafie2}, the authors in~\cite{Wu2} develop a statistical framework for the indoor THz channel. The developed framework accounts for the line-of-sight (LOS) and NLOS THz communications in indoor environments and approximates the fading distribution from the multi-ray THz channel model. 

Despite the coverage limitations of THz communications, modeling and analyzing hybrid RF/THz networks to satisfy coverage and high rate requirements is not yet thoroughly investigated. Most of the works in the literature are focused on assessing the performance of THz-only networks while characterizing THz propagation accurately. The work in~\cite{Sayehvand} is one of the few exceptions that considers a coexisting sub-$6$~GHz and dense THz wireless network. However, this work only focuses on a 2D environment and does not account for the impact of directional antennas and blockages, which can significantly affect the THz transmission. In addition, the infinite Poisson point process (PPP) is used to model the THz network in~\cite{Sayehvand}, which does not fit realistic indoor uses cases of THz deployment. The work in~\cite{Shi} considers a THz-only network while accounting for the finite nature of the THz network and evaluates the performance of both central users and edge users. Another heterogeneous network consisting of macro BSs that operate at sub-$6$~GHz, unmanned aerial vehicles (UAVs) that operate at mmWave frequencies, and small BSs using both mmWave and THz communications is proposed in~\cite{Raja}. However, this work captures the impact of blockages on mmWave communications only and ignores it for THz communications. To the best of the authors' knowledge, none of the previous research works presented a comprehensive analytical framework to characterize coverage probability and rate in a coexisting RF and THz finite indoor network. As a result, this work aims to address the details of this problem using stochastic geometry and to devise useful recommendations for THz deployment.
\vspace{-0.3cm}
\subsection{Contributions}
This paper considers a hybrid RF/THz network, where THz and RF APs coexist to provide coverage and throughput for UEs in an indoor open office environment. The THz APs use directional antennas to cope with high loss levels and limited coverage of THz communications and are affected by existing blockages in the environment and beam- steering errors of directional antennas. Using tools from stochastic geometry, we aim to assess the performance of the coexisting RF and THz network, highlighting the impact of different system parameters. To this extent, the main contributions of this paper can be described as follows:
\begin{itemize}
	\item We consider an open office indoor environment where a finite number of RF and THz APs coexist to provide users coverage and rate. We model the APs network as a binomial point process (BPP), fitting more realistic indoor applications than the infinite PPP used in most literature works. The developed model accounts for the molecular absorption loss, which significantly affects the THz propagation. Furthermore, an accurate analytical model is used to account for the human blockages in the environment, in addition to the directional antennas at both the THz APs and users and the encountered beam-steering errors.
	\item We devise a tractable analytical framework, using tools from stochastic geometry, to characterize the coverage probability and average achievable rate of the coexisting RF and THz network. Specifically, we derive the association probabilities with an RF AP and a THz AP and the conditional coverage probabilities and average rates. Finally, we use the law of total probability to determine the considered metrics. Unlike most literature, the proposed analytical framework provides exact expressions for the coverage probability and average rate rather than relying on approximations. The analytical results are validated using Monte-Carlo simulations.
	\item Based on the developed framework, we study how different system parameters affect the network performance. Such parameters include the fraction of THz APs and the total number of APs, the THz bias term, the location of the UE in the finite area, and the beam-steering errors. The obtained results capture the coverage/rate trade-off imposed by densifying the network with THz APs and devise useful design guidelines for THz deployment.
\end{itemize}
\begin{table}[t!]
	\centering
	\caption{Notations Summary.}
	\vspace{-0.3cm}
		\begin{tabular}{m{0.18\linewidth}|m{0.77\linewidth}}
		\hline\hline
		\textbf{Notation} & \textbf{Description} \\\hline
	    $r_d$ & Radius of the disk in which the APs are distributed\\\hline
	    $h_{\mathrm{A}}$ & Height of the APs with reference to the ground level\\\hline
		$N_{\mathrm{A}}$ & Number of APs \\\hline
		$v_0$, $h_{\mathrm{U}}$ & Distance from the UE to origin, height of UE from the ground level \\\hline
		$\Phi_{\mathrm{A}}$, $\Phi_{\mathrm{R}}$, $\Phi_{\mathrm{T}}$, $\Phi_{\mathrm{L}}$, $\Phi_{\mathrm{N}}$ &Set of APs, RF APs, THz APs, LOS THz APs, or NLOS THz APs, respectively \\\hline
		$\delta_{\mathrm{T}}$  & Fraction of THz APs \\\hline
		$P_{\mathrm{L}}(\cdot)$, $P_{\mathrm{N}}(\cdot)$ & Probability of the THz AP having a LOS or a NLOS connection with the UE, respectively \\\hline
		$P_{\mathrm{R}}$, $P_{\mathrm{T}}$ & Transmit power of RF and THz APs, respectively\\\hline
		$W_{\mathrm{R}}$, $W_{\mathrm{T}}$ & RF and THz bandwidths \\\hline
		$f_{\mathrm{R}}$, $f_{\mathrm{T}}$& RF frequency and THz frequency, respectively\\\hline
		$k_a(f_{\mathrm{T}})$ & THz absorption coefficient \\\hline
		$\sigma_{R}^2$, $\sigma_{T}^2$ & Noise power  of RF and THz communications, respectively\\\hline	
		$\alpha_{\mathrm{R}}$, $\alpha_{\mathrm{L}}$, $\alpha_{\mathrm{N}}$ & Path-loss exponent parameter for RF AP, LOS THz AP, or NLOS THz AP, respectively \\\hline
		$m_{\mathrm{L}}$, $m_{\mathrm{N}}$ & Nakagami-m fading parameter for RF AP, LOS THz AP, or NLOS THz AP, respectively \\\hline
		$G_{\mathrm{T}}^{\text{max}}$, $G_{\mathrm{T}}^{\text{min}}$, $\varphi_{\mathrm{T}}$ & Antenna parameters for THz AP \\\hline
		$G_{\mathrm{U}}^{\text{max}}$, $G_{\mathrm{U}}^{\text{min}}$, $\varphi_{\mathrm{U}}$ & Antenna parameters for UE \\\hline
		$\lambda_{B}$, $r_{B}$, $h_{B}$& Density, radius and height of blockages\\\hline
		$\sigma_{\varepsilon_{\mathrm{T}}}$, $\sigma_{\varepsilon_{\mathrm{U}}}$ & THz/UE beam-steering error \\\hline
		$\chi_{\mathrm{R},x_{i}}$, $\chi_{\mathrm{L},x_{i}}$, $\chi_{\mathrm{N},x_{i}}$ & Small scale fading gain between the UE and an RF AP, a LOS THz AP or a NLOS THz AP, respectively, located at $\textbf{x}_{i}$\\\hline
		$d_{\mathrm{R},x_{i}}$, $d_{\mathrm{L},x_{i}}$, $d_{\mathrm{N},x_{i}}$ & Distance between UE and an RF AP, a LOS THz or a NLOS THz AP located at $\textbf{x}_{i}$, respectively\\\hline
		$d_{\mathrm{R}}$, $d_{\mathrm{L}}$, $d_{\mathrm{N}}$ & Distance between the UE and its nearest
    	RF AP, LOS THz AP or NLOS THz AP, respectively\\\hline
		$x_{\mathrm{R}}$, $x_{\mathrm{L}}$, $x_{\mathrm{N}}$ & Distance between the UE and its serving AP assuming that the UE is associating to an RF AP, a LOS THz AP or a NLOS THz AP, respectively\\\hline
		$A_{\mathrm{R}}$, $A_{\mathrm{L}}$, $A_{\mathrm{N}}$ & Probability of association between the UE and RF AP, LOS THz AP, or NLOS THz AP, respectively \\\hline
		$P_{cov,\mathrm{R}}$, $P_{cov,\mathrm{L}}$, $P_{cov,\mathrm{N}}$ & Conditional coverage probability given that the UE is associated with an RF AP, a LOS THz AP, or a NLOS THz AP, respectively\\\hline
		$\tau_{\mathrm{R}}$, $\tau_{\mathrm{L}}$, $\tau_{\mathrm{N}}$ & Conditional average rate given that the UE is associated with an RF AP, a LOS THz AP, or a NLOS THz AP, respectively\\\hline
		$P_{cov}$, $\tau$ & Overall coverage probability and overall coverage rate\\\hline
		$\theta$ & $\mathrm{SINR}$ threshold \\\hline\hline
	\end{tabular}
	\label{table:parameters}
	\vspace{-0.5cm}
\end{table}
\vspace{-0.5cm}
\subsection{Organization and Notations}
Throughout the paper, the subscripts $\{\cdot\}_{\mathrm{A}}$ and $\{\cdot\}_{\mathrm{U}}$ refer to AP and user, respectively. The subscripts $\{\cdot\}_{\mathrm{T}}$ and $\{\cdot\}_{\mathrm{R}}$ indicate THz and RF communications. The subscripts $\{\cdot\}_{\mathrm{L}}$, $\{\cdot\}_{\mathrm{N}}$ differentiate between THz LOS and THz NLOS, respectively. The symbol $\mathbb{P}\{\cdot\}$ refers to probability, while $\mathbb{E}[\cdot]$, $\mathcal{W}[\cdot]$ and $\mathcal{L}_x(\cdot)$ denote the expectation, the Lambert W-function defined as the inverse function of $f(w)=we^w$, and the Laplace transform of a random variable $x$, respectively. The rest of notations are presented in Table~\ref{table:parameters}.

The remainder of the paper is organized as follows. We describe the RF and THz coexisting system model in Section~\ref{sec:system_model}. Section~\ref{sec:association_serving} presents the derivations of the association probabilities and serving distance distributions. Section~\ref{sec:coverage_rate} describes the derivations related to the coverage probability and the average achievable rate. Numerical results are discussed in Section~\ref{sec:results} and validated using Monte-Carlo simulations. Finally, we conclude the paper in Section~\ref{sec:conclusion}.
\begin{figure}[t]
	\centering
	\includegraphics[width=0.5\linewidth]{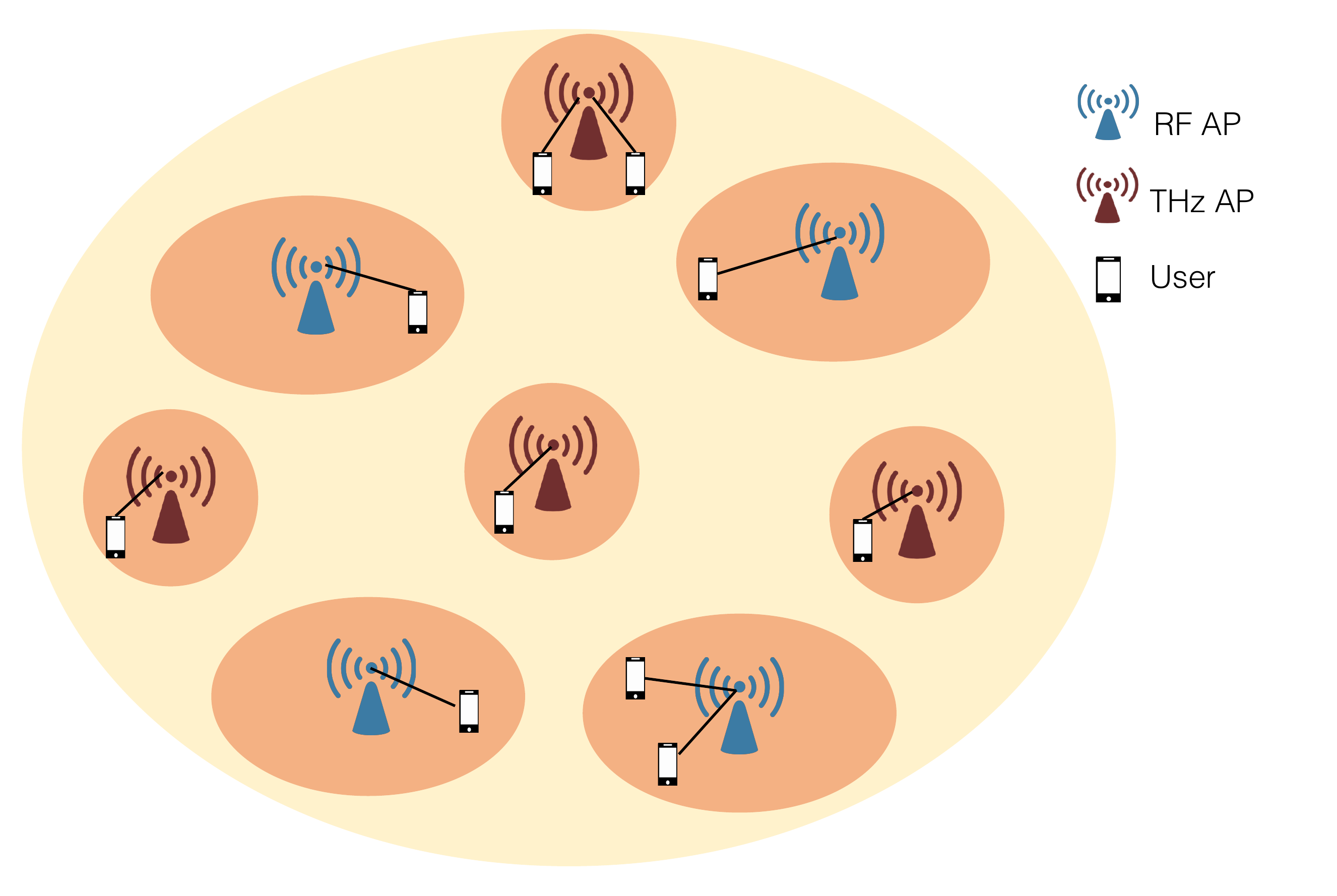}
	\vspace{-0.5cm}
	\caption{Network architecture of the RF and THz coexisting network.}	
	\label{fig:model2}
	\vspace{-0.5cm}
\end{figure}
\vspace{-0.5cm}
\section{System model}\label{sec:system_model}
\subsection{Network Model}\label{sec:network_model}
In this work, we consider a downlink (DL) wireless network where a fixed number $N_{\mathrm{A}}$ of RF and THz access points (APs) are mounted on the ceiling of an indoor finite area $\mathcal{A}$ to serve the user equipments (UEs) as shown in Fig.~\ref{fig:model2}. Thus, the locations of the APs are modeled as a uniform binomial point process (BPP) $\Phi_{\mathrm{A}}\triangleq \{\textbf{x}_i\}$, where $\textbf{x}_i$ refers to the location of the $i$-th AP in the finite region $\mathcal{A}=\textbf{b}(\textbf{o'},r_d)$ modeled as a disk of radius $r_d$ centered around $\textbf{o'} = (0, 0, h_{\mathrm{A}})$, where $h_A$ is the height of the APs from the ground level. We consider the performance of a reference UE located at a fixed height $h_{\mathrm{U}}$ from the ground level and at an arbitrary location $\textbf{v}_0$ from the origin $\textbf{o} = (0, 0, 0)$ assumed to be in the same UE's plane. As the BPP remains invariant with respect to the orientation of the axes, we can consider, without loss of generality, that the UE is located on the x-axis, i.e., the location of the UE is $\textbf{v}_0=(v_0,0,0)$, where $||\textbf{v}_{0}||=v_0 \in [0,r_d]$ and $||\cdot||$ is the euclidean norm. Furthermore, the APs are deployed at a height of $h_{\mathrm{A}}-h_{\mathrm{U}}$ from the origin plane. A fraction $\delta_{T}$ of the APs are THz APs that transmit with the same power $P_{\mathrm{T}}$ while $(1-\delta_{\mathrm{T}})$ are RF APs transmitting with a power $P_{\mathrm{R}}$. From the UE's point of view, the set of APs is decomposed into two independent BPPs, i.e., $\Phi_a=\Phi_{\mathrm{T}}\cup\Phi_{\mathrm{R}}$ where $\Phi_{\mathrm{T}}$ and $\Phi_{\mathrm{R}}$ denote the sets of THz and RF APs. Note here that $\delta_{\mathrm{T}}$ should be chosen such that $\delta_{\mathrm{T}}N_{\mathrm{A}}$, which represents the total number of THz APs, is always an integer number. However, the obtained analytical derivations are still applicable $\forall \delta_{\mathrm{T}}\leq 1$. The UE evaluates the quality of the channel from each existing AP and associates to a specific AP according to the association rule. For the RF communication, the RF APs use omni-directional antennas to communicate with single antenna UEs. However, THz APs are equipped with dedicated antenna arrays that operate on the considered THz frequency to serve UEs with directional antennas.
\vspace{-0.5cm}
\subsection{RF Channel Model}\label{sec:RF_channel_model}
The RF communication is affected by a distance dependent large scale fading and a small scale Rayleigh fading that follows the exponential distribution with unit mean. Thus, the received power $P^{r}_{\mathrm{R},x_i}$ from the $i$-th RF AP located at $\textbf{x}_i$ is given by $P^{r}_{\mathrm{R},x_i}=P_{\mathrm{R}}\gamma_{\mathrm{R}}d_{\mathrm{R},x_i}^{-\alpha_\mathrm{R}}\chi_{\mathrm{R},x_i}$, where $P_\mathrm{R}$ is the RF AP transmit power, $\gamma_{\mathrm{R}}=\frac{c^2}{(4\pi f_{\mathrm{R}})^2}$, $d_{\mathrm{R},x_i}=||\textbf{x}_{i}-\textbf{v}_{0}||$ is the distance from the UE to the $i$-th RF AP, $\alpha_{\mathrm{R}}$ is the path loss exponent, $\chi_{\mathrm{R},x_i}$ is the gain of the small scale Rayleigh fading, $f_{\mathrm{R}}$ is the carrier frequency of the RF communication and $c=3\times 10^8$~m/s is the speed of light. 
\begin{figure}[t]
	\centering
	\begin{minipage}[t]{0.47\linewidth}
		\includegraphics[width=\linewidth]{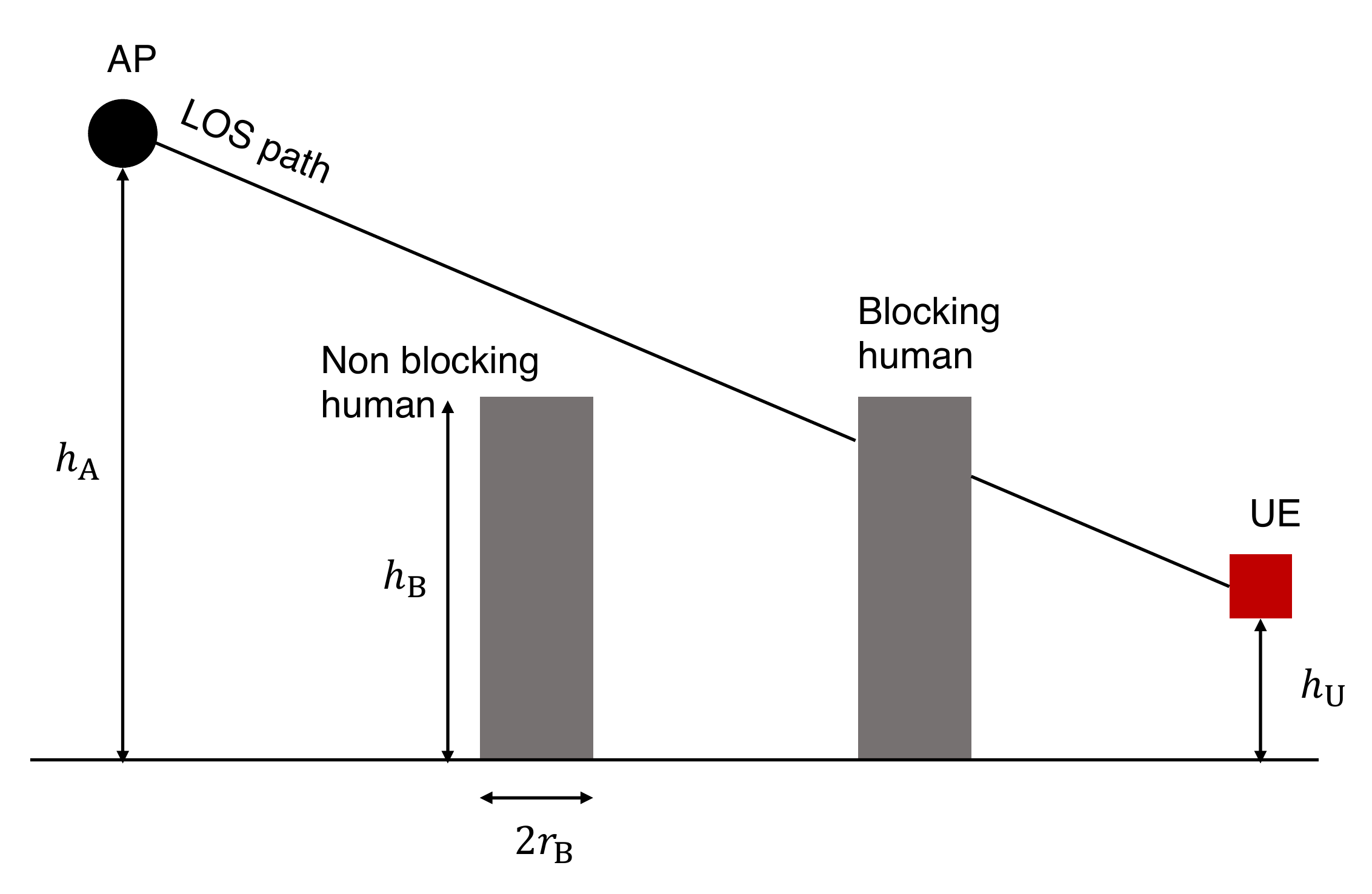}
		\vspace{-0.8cm}
		\caption{Vertical View of the human-blocking scenario for an AP-UE link.}	
		\label{fig:blockage_model}
	\end{minipage}
	\begin{minipage}[t]{0.47\linewidth}
		\includegraphics[width=\linewidth,height=0.68\linewidth]{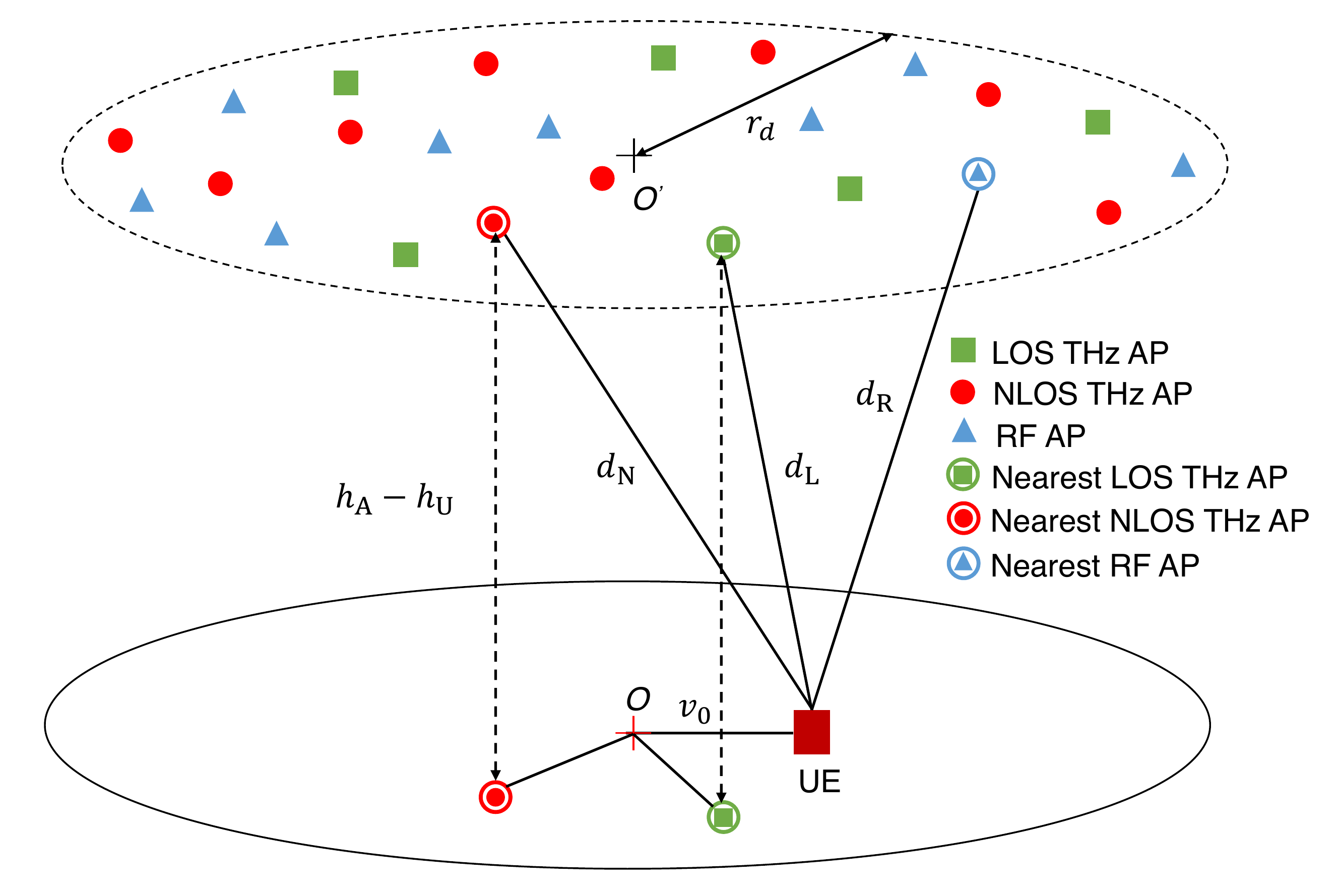}
		\vspace{-0.8cm}
		\caption{An example of the BPP distribution of APs in the disk $\mathcal{A}$. The set of APs is divided into RF, LOS THz, and NLOS THz APs.}	
		\label{fig:model}
	\end{minipage}
	\vspace{-0.6cm}
\end{figure}
\vspace{-0.5cm}
\subsection{THz Channel Model}\label{sec:THz_channel_model}
\subsubsection{Blockage Model}
The THz propagation is highly affected by the obstacles in the environment. The presence of these obstacles breaks the LOS connection and converts it to a NLOS connection. In this work, we study an open office environment as defined in the 3GPP standards~\cite{3gpp38900}, where only human blockers exist and might affect the wireless connection. Fig.~\ref{fig:blockage_model} shows the case when a human body blocks the AP-UE LOS link and convert it to a NLOS link. Note that we will consider a more sophisticated blockage model that accounts for indoor specific obstacles such as walls and furniture in a future extension of this work. We assume that human blockages are modeled as a random circle process of radius $r_{\mathrm{B}}$ and height $h_{\mathrm{B}}$. Specifically, the bottom center of the cylinder characterizing a human blocker is modeled as a 2D homogeneous PPP of density $\lambda_{\mathrm{B}}$. As a consequence of the high penetration loss of THz communication, if the LOS link is blocked, the UE can only communicate with the serving AP through reflected NLOS links. The probability of having a LOS connection between a THz AP located at a distance $r$ from the UE, denoted $\kappa_{\mathrm{L}}(r)$ is calculated as the null probability of the human blockages PPP. Defining $\beta=2\lambda_{\mathrm{B}}r_{\mathrm{B}}\frac{|h_{\mathrm{B}}-h_{\mathrm{U}}|}{|h_{\mathrm{A}}-h_{\mathrm{U}}|}$, where $h_{\mathrm{A}}$ and $h_{\mathrm{U}}$ are the corresponding heights of the AP and the UE, the LOS and NLOS probabilities are given in~\cite{Wu} as
\begin{equation}
	\kappa_{\mathrm{L}}(r)=e^{-\beta \sqrt{r^2-(h_{\mathrm{A}}-h_{\mathrm{U}})^2}},
	\label{eq:PL}
	\vspace{-0.5cm}
\end{equation} 
\begin{equation}
\kappa_{\mathrm{N}}(r)=1-e^{-\beta \sqrt{r^2-(h_{\mathrm{A}}-h_{\mathrm{U}})^2}},
\label{eq:PN}
\end{equation}   
where $\sqrt{r^2-(h_{\mathrm{A}}-h_{\mathrm{U}})^2}$ is the Euclidean horizontal distance that separate the reference UE from the projection of the AP location on the UE plane. Based on the considered channel model, the UE is exposed to either a LOS or a NLOS connection with any THz AP. Although this assumption becomes unrealistic in circumstances where APs are very close to each other and are likely to face the same LOS or NLOS situations, it can simplify the mathematical analysis significantly and leads as result to tractable analytical expressions. From the UE's perspective, the set of THz APs $\Phi_{\mathrm{T}}$ will be divided into two subsets, i.e, $\Phi_{\mathrm{T}}=\Phi_{\mathrm{L}}\cup\Phi_{\mathrm{N}}$, where $\Phi_{\mathrm{L}}$ and $\Phi_{\mathrm{N}}$ represent the set of THz APs which are in LOS or NLOS situations with the reference UE, respectively. This decomposition is done by mapping each point of $\Phi_{\mathrm{T}}$ into one of the disjoint sets $\Phi_{\mathrm{L}}$ and $\Phi_{\mathrm{N}}$ with probabilities $\kappa_{\mathrm{L}}(r)$ and $\kappa_{\mathrm{N}}(r)$, respectively, where $r$ represents the location of the corresponding AP. Fig.~\ref{fig:model} shows a realization of the BPP of APs with the three subsets for RF, LOS THz and NLOS THz APs.
\subsubsection{Propagation Model}
The THz communication is highly affected by the molecular absorption loss caused by the existing water molecules in the atmosphere. Thus, the large scale fading is modeled as a deterministic exponent power loss propagation model. Furthermore, since the THz communications are very susceptible to the availability of LOS paths, the Rayleigh fading assumption is invalid and the small scale fading follows a Nakagami-m distribution. We use different path loss exponents and Nakagami-m parameters for THz LOS and NLOS transmissions ($\alpha_{\mathrm{L}}$ and $m_{\mathrm{L}}$ for LOS links and $\alpha_{\mathrm{N}}$ and $m_{\mathrm{N}}$ for NLOS links). Thus, the channel fading gain $\chi_{\xi,x_i}$, $\xi\in\{\mathrm{L},\mathrm{N}\}$, between the $i$-th THz AP located at $\textbf{x}_i$ and the UE, follows the Gamma distribution with shape and scale parameters given by $\big(m_{\xi},\frac{1}{m_{\xi}}\big)$ and with a complementary cumulative distribution function (CCDF) given by
\begin{equation}
	\bar{F}_{\chi_{\xi,x_i}}(x)=\sum_{k=0}^{m_{\xi}-1}\frac{\left(m_{\xi}x\right)^{k}}{k!}\exp\left(-m_{\xi}x\right).
\end{equation}

Considering probabilistic LOS and NLOS transmissions, the path loss between the reference UE and a THz AP can be expressed as $l_{\xi}(z)=\gamma_{\mathrm{T}}e^{-k_a(f_{\mathrm{T}})z}z^{-\alpha_{\xi
}}$, where $\gamma_{\mathrm{T}}=\frac{c^2}{(4\pi f_T)^2}$, $\xi\in\{\mathrm{L},\mathrm{N}\}$ specifies if the THz AP has a LOS connection or NLOS connection with the UE, $k_a(f_{\mathrm{T}})$ is the molecular absorption coefficient, $f_{\mathrm{T}}$ is the frequency of operation of the THz communication and $z$ is the distance separating the UE from the considered AP.
\subsubsection{Antenna Model}
To overcome large path and absorption losses, directional antennas are usually used in the THz frequencies because of the small antenna sizes which brought great potential for large multiple input multiple output (MIMO) arrays implementations \cite{sarieddeen2019terahertz,sarieddeen2021overview}. To approximate the array patterns of the THz APs and the corresponding UE, we use the antenna model given below
\vspace{-0.5cm}
\begin{equation}
	G_s(\varphi)=\begin{cases}
	G_s^{(\mathrm{max})}, & |\varphi|\leq\varphi_s\\
	G_s^{(\mathrm{min})}, & |\varphi|>\varphi_s
	\end{cases},
	\label{eq:antenna_gain}
\end{equation}
where $\varphi\in[-\pi,\pi)$ represents the angle of boresight direction, $G_s^{(\mathrm{max})}$, $G_s^{(\mathrm{min})}$, and $\varphi_s$ denote the main and side lobes gains and the beamwidth of the THz APs antennas and the UEs antennas operating in the THz band ($s\in\{\mathrm{T},\mathrm{U}\}$), respectively. 

When the UE chooses to associate with a THz AP, both UE and AP steer their directional antennas so as to maximize the directionality gain. In the absence of beam-steering errors, the UE can benefit from the gains of the main lobes of its antenna and the antenna of the THz AP. The directionality gain on the desired link can therefore be expressed as $G_{\mathrm{T},0}=G_{\mathrm{T}}^{\mathrm{max}}G_{\mathrm{U}}^{\mathrm{max}}$. However, achieving perfect alignment is not always feasible as it requires extremely narrow beams and further processing at both sides. To account for the beam-steering error on the THz-UE connection, we used the model proposed in~\cite{Wildman}. Thus, we denote by $\varepsilon_{\mathrm{T}}$ and $\varepsilon_{\mathrm{U}}$ the added beam-steering errors on the THz AP and the UE respectively. To this extend, we assume that $\varepsilon_{\mathrm{T}}$ and $\varepsilon_{\mathrm{U}}$ can be modeled using Gaussian distribution with zero mean and variances $\sigma_{\varepsilon_{\mathrm{T}}}^2$ and $\sigma_{\varepsilon_{\mathrm{U}}}^2$, respectively. Furthermore, $\varepsilon_{\mathrm{T}}$ and $\varepsilon_{\mathrm{U}}$ are independent of each other and are symmetrically distributed around the error-free beam-steering angles. Thus, $|\varepsilon_{s}|$, for $s \in \{\mathrm{T},\mathrm{U}\}$, follows a half-normal distribution with a cumulative distribution function (CDF) $F_{\lvert\varepsilon_{s}\rvert}(x)=\mathrm{erf}\left(\frac{x}{\sqrt{2}\sigma_{\varepsilon_{s}}}\right)$, where $\mathrm{erf}(\cdot)$ is the error function.

As the antenna gain of the THz AP and the UE is a discrete random variable that can takes only two values as given in (\ref{eq:antenna_gain}), the corresponding probability mass function (PMF) in the presence of beam-steering errors can be expressed as 
\begin{equation}
\begin{aligned}
f_{G_s}(g)&=F_{|\varepsilon_{\mathrm{s}}|}\left(\frac{\varphi_s}{2}\right)\delta\left(g-G_s^{\text{max}}\right)+\bar{F}_{|\varepsilon_{s}|}\left(\frac{\varphi_s}{2}\right)\delta\left(g-G_s^{\text{min}}\right),
\end{aligned}
\label{eq:fG0}
\end{equation}
where $\delta(\cdot)$ is the Dirac delta function, $\bar{F}_{|\varepsilon_{s}|}(x)=\left(1-F_{|\varepsilon_{s}|}(x)\right)$, $G_s^{\text{max}}$, $G_s^{\text{min}}$ and $\varphi_s$ are gains of the the main and side lobes and beamwidth for $s \in \{\mathrm{T},\mathrm{U}\}$. Accordingly, the directionality gain on the desired link $G_{\mathrm{T},0}$ is also a discrete random variable that can take the values $G_k$ with probabilities $p_{k,0}$, ($k\in\{1,2,3,4\}$) as given in Table~\ref{table:antenna}.

As the UE associates with a THz AP, the remaining THz APs will act as interferers that can affect the THz connection. However, the antennas of the interfering THz APs are not necessarily steered towards the reference UE that can receive interference from either the main lobe or the side lobe of the directional antenna. To account for the THz interference, we consider that the steering angles between the $i$-th THz AP located at $\textbf{x}_i$ and the UE are uniformly distributed in $[0,2\pi]$. Thus, the directionality gain $G_{\mathrm{T},x_{i}}$ is a discrete random variable that can take four different values $G_k$, ($k\in\{1,2,3,4\}$) with probabilities $p_{k}$ as given in Table~\ref{table:antenna}, where $v_{\mathrm{T}}=\frac{\varphi_{\mathrm{T}}}{2\pi}$ and $v_{\mathrm{U}}=\frac{\varphi_{\mathrm{U}}}{2\pi}$, $\varphi_{\mathrm{T}}$ and $\varphi_{\mathrm{U}}$ are the the beamwidth for the THz APs antennas and the UEs antennas. The received power at the reference UE from the $i$-th THz AP placed at $\textbf{x}_i$ is given as $P^{r}_{\xi,x_i}=P_{\mathrm{T}}\gamma_{\mathrm{T}}G_{\mathrm{T},x_{i}}e^{-k_a(f_{\mathrm{T}})d_{\xi,x_i}}d_{\xi,x_i}^{-\alpha_{\xi}}\chi_{\xi,x_i}$, where $\xi\in\{\mathrm{L},\mathrm{N}\}$ indicates if the THz AP located at $\textbf{x}_i$ has a LOS or a NLOS connection with the reference UE. $\chi_{\xi,x_i}$ is the small scale Nakagami fading, $\gamma_{\mathrm{T}}=\frac{c^2}{(4\pi f_{\mathrm{T}})^2}$, $P_\mathrm{T}$ is the THz AP transmit power, $d_{\xi,x_i}=||\textbf{x}_{i}-\textbf{v}_{0}||$ is the distance from the UE to the $i$-th THz AP, $\alpha_{\xi}$ is the path loss exponent, $f_{\mathrm{T}}$ is the THz carrier frequency. 
\begin{table}[t]
	\centering
	\caption{Probability mass function.}
	\vspace{-0.3cm}
	\begin{tabular}{c|c|c|c|c}
		\hline
		\textbf{k} & \textbf{1} & \textbf{2} & \textbf{3} & \textbf{4} \\\hline
		 $G_{k}$ & $G_{\mathrm{T}}^{\mathrm{max}}G_{\mathrm{U}}^{\mathrm{max}}$&$G_{\mathrm{T}}^{\mathrm{max}}G_{\mathrm{U}}^{\mathrm{min}}$ & $G_{\mathrm{T}}^{\mathrm{min}}G_{\mathrm{U}}^{\mathrm{max}}$ & $G_{\mathrm{T}}^{\mathrm{min}}G_{\mathrm{U}}^{\mathrm{min}}$\\\hline
		 $p_{k,0}$ & $F_{\lvert\varepsilon_{\mathrm{T}}\rvert}(\frac{\varphi_{\mathrm{T}}^{*}}{2})F_{\lvert\varepsilon_{\mathrm{U}}\rvert}(\frac{\varphi_{\mathrm{U}}^{*}}{2})$  & $F_{\lvert\varepsilon_{\mathrm{T}}\rvert}(\frac{\varphi_{\mathrm{T}}^{*}}{2})\bar{F}_{\lvert\varepsilon_{\mathrm{U}}\rvert}(\frac{\varphi_{\mathrm{U}}^{*}}{2})$ & $\bar{F}_{\lvert\varepsilon_{\mathrm{T}}\rvert}(\frac{\varphi_{\mathrm{T}}^{*}}{2})F_{\lvert\varepsilon_{\mathrm{U}}\rvert}(\frac{\varphi_{\mathrm{U}}^{*}}{2})$ & $\bar{F}_{\lvert\varepsilon_{\mathrm{T}}\rvert}(\frac{\varphi_{\mathrm{T}}^{*}}{2})\bar{F}_{\lvert\varepsilon_{\mathrm{U}}\rvert}(\frac{\varphi_{\mathrm{U}}^{*}}{2})$\small\\\hline
		 $p_{k}$ & $v_{\mathrm{T}} v_{\mathrm{U}}$ & $v_{\mathrm{T}}(1-v_{\mathrm{U}})$& $(1-v_{\mathrm{T}})v_{\mathrm{U}}$ & $(1-v_{\mathrm{T}})(1-v_{\mathrm{U}})$ \\\hline
	\end{tabular}
	\label{table:antenna}
	\vspace{-0.5cm}
\end{table}
\vspace{-0.5cm}
\subsection{Association Policy and SINR}
In this paper, we assume that the association decision is taken by referring to long term evaluation of the channel instead of short term metrics. Thus, the UE associates with the AP that has the strongest average biased received power (BRSP). Furthermore, biased association is considered to avoid under-utilization of THz APs. It is worth mentioning that the UE has three association options: an RF AP, a LOS THz AP or a NLOS THz AP depending on the averaged biased received power. Note here that the nearest AP is not necessarily the AP that provides the strongest received power because of the difference in path-loss parameters, transmit powers and antenna configurations. However, within a specific set of APs, i.e., within each set of RF, LOS THz and NLOS THz APs, the aforementioned parameters are the same for all connections. Thus, for a particular set, the nearest AP has a larger average received power than that provided by all the remaining APs in this set. As result, the serving AP is always the closest RF, LOS THz or NLOS THz AP. According to the considered association rule and the assumptions that $\mathbb{E}\left[\chi_{\mathrm{R},x_i}\right]=\mathbb{E}\left[\chi_{\mathrm{L},x_i}\right]=\mathbb{E}\left[\chi_{\mathrm{N},x_i}\right]=1$, the serving AP is given as $\mathrm{argmax}\{P_{\mathrm{R}}\gamma_{\mathrm{R}}d_{\mathrm{R}}^{-\alpha_{\mathrm{R}}},B_{\mathrm{T}}P_{\mathrm{T}}\gamma_{\mathrm{T}}G_{\mathrm{T},0}^{(\mathrm{mean})}e^{-k_{a}(f_{\mathrm{T}})d_{\mathrm{L}}}d_{\mathrm{L}}^{-\alpha_{\mathrm{L}}},B_{\mathrm{T}}P_{\mathrm{T}}\gamma_{\mathrm{T}}G_{\mathrm{T},0}^{(\mathrm{mean})}e^{-k_{a}(f_{\mathrm{T}})d_{\mathrm{N}}}d_{\mathrm{N}}^{-\alpha_{\mathrm{N}}}\}$, where $\gamma_T=\frac{c^2}{(4\pi f_T)^2}$, $d_{\mathrm{R}}=\min\limits_{\forall \textbf{x}_i\in\Phi_{\mathrm{R}}}d_{\mathrm{R},x_{i}}$, $d_{\mathrm{L}}=\min\limits_{\forall \textbf{x}_i\in\Phi_{\mathrm{L}}}d_{\mathrm{L},x_{i}}$, and $d_{\mathrm{N}}=\min\limits_{\forall \textbf{x}_i\in\Phi_{\mathrm{N}}}d_{\mathrm{N},x_{i}}$ are the distances from the UE to the nearest RF, LOS THz and NLOS THz APs. $G_{\mathrm{T},0}^{(\mathrm{mean})}=\sum\limits_{k=1}^{4}p_{k,0}G_{k}$ is the average directionality gain on the desired link, where $p_{k,0}$ and $G_{k}$ are given in Table~\ref{table:antenna}. $B_{\mathrm{T}}$ is the THz bias parameter. For $B_{\mathrm{T}}>1$, the UE is encouraged to associate more with THz APs. For $0\leq B_{\mathrm{T}}<1$, the association with RF APs is encouraged and when $B_{\mathrm{T}}=0$, the association is taken based on the average reference signal received power (RSRP) by the UE.

As the objective of this work is to assess the DL performance of a finite coexisting RF and THz network, the main performance metrics used are the DL coverage probability and the average rate. To this extend, we define the coverage probability, denoted as $P_{cov}$, as the probability that the $\mathrm{SINR}$ at the reference UE exceeds a threshold $\theta$. When the UE associates with an RF AP, the $\mathrm{SINR}$ can be formulated as:
\begin{equation}
	\mathrm{SINR}_{\mathrm{R}}=\frac{P_{\mathrm{R}} \gamma_{\mathrm{R}} x_{\mathrm{R}}^{-\alpha_{\mathrm{R}}}\chi_{\mathrm{R},0}}{I_{\mathrm{R}}+\sigma_{\mathrm{R}}^2},
	\label{eq:SINR_R}
\end{equation}
where $\chi_{\mathrm{R},0}$ is the small scale fading experienced by the UE on the desired link, $x_{\mathrm{R}}$ is the distance separating the UE from its serving RF AP, $\sigma_{\mathrm{R}}^2$ is the average noise and $I_{\mathrm{R}}$ is the interference at the reference UE from the interfering RF APs and is given by
\begin{equation}
	I_{\mathrm{R}}=\sum\limits_{\textbf{x}_{i}\in\Phi_{\mathrm{R}}/x_{\mathrm{R}}}P_{\mathrm{R}}\gamma_{\mathrm{R}} d_{\mathrm{R},x_{i}}^{-\alpha_{\mathrm{R}}}\chi_{\mathrm{R},x_{i}},
	\label{eq:IR}
\end{equation}
where $d_{\mathrm{R},x_{i}}$ is the distance separating the RF AP located at $\textbf{x}_i$ from the reference UE. 

Similarly, the $\mathrm{SINR}$ of the reference UE when associating with a LOS THz AP is given as
\begin{equation}
\mathrm{SINR}_{\mathrm{L}}=\frac{P_T \gamma_T G_{\mathrm{T},0} e^{-k_a(f_{\mathrm{T}})x_{\mathrm{L}}}x_{\mathrm{L}}^{-\alpha_{\mathrm{L}}}\chi_{\mathrm{L},0}}{I_{\mathrm{L}}+\sigma_{\mathrm{T}}^2},
\label{eq:SINR_L}
\end{equation}
where $x_{\mathrm{T}}$ is the distance that separates the UE from its serving LOS THz AP, $G_{\mathrm{T},0}$ is the directionality gain which has a PMF given in Table~\ref{table:antenna}, $\chi_{\mathrm{L},0}$ denotes the small scale fading, $\sigma_T^2$ is the thermal noise at the UE, $k_a(f)$ is the molecular absorption coefficient, $I_{\mathrm{L}}$ is the aggregate interference from THz APs that can have LOS links with the reference UE except the serving LOS THz AP and the THz APs with NLOS links. Thus, $I_{\mathrm{L}}$ is given as
\begin{equation}
I_{\mathrm{L}}=\sum_{\textbf{x}_{i}\in\Phi_{\mathrm{L}}/x_{\mathrm{L}}}P_{\mathrm{T}}\gamma_{\mathrm{T}}G_{\mathrm{T},x_{i}}e^{-k_a(f_{\mathrm{T}})d_{\mathrm{L},x_{i}}}d_{\mathrm{L},x_{i}}^{-\alpha_{\mathrm{L}}}\chi_{\mathrm{L},x_{i}}+\sum_{\textbf{x}_{i}\in\Phi_{\mathrm{N}}}P_{\mathrm{T}}\gamma_{\mathrm{T}}G_{\mathrm{T},x_{i}}e^{-k_a(f_{\mathrm{T}})d_{\mathrm{N},x_{i}}}d_{\mathrm{N},x_{i}}^{-\alpha_{\mathrm{N}}}\chi_{\mathrm{N},x_{i}},
\label{eq:IL}
\end{equation}
where $G_{\mathrm{T},x_i}$ is the directionality gain between the $i$-th interfering THz AP and the reference UE, $d_{\mathrm{L},x_i}$ and $d_{\mathrm{N},x_i}$ denote the distances between the THz AP located at $\textbf{x}_i$ and the reference UE for LOS and NLOS transmission, respectively.

Finally, the $\mathrm{SINR}$ of the reference UE when associating with a NLOS THz AP is given as
\begin{equation}
\mathrm{SINR}_{\mathrm{N}}=\frac{P_T \gamma_T G_{\mathrm{T},0} e^{-k_a(f_{\mathrm{T}})x_{\mathrm{N}}}x_{\mathrm{N}}^{-\alpha_{\mathrm{N}}}\chi_{\mathrm{N},0}}{I_{\mathrm{N}}+\sigma_{\mathrm{T}}^2},
\label{eq:SINR_N}
\end{equation}
The interference $I_{\mathrm{N}}$ from the NLOS THz APs except the serving AP and from all LOS THz APs is given as
\begin{equation}
I_{\mathrm{N}}=\sum_{\textbf{x}_{i}\in\Phi_{\mathrm{N}}/x_{\mathrm{N}}}P_{\mathrm{T}}\gamma_{\mathrm{T}}G_{\mathrm{T},x_{i}}e^{-k_a(f_{\mathrm{T}})d_{\mathrm{N},x_{i}}}d_{\mathrm{N},x_{i}}^{-\alpha_{\mathrm{N}}}\chi_{\mathrm{N},x_{i}}+\sum_{\textbf{x}_{i}\in\Phi_{\mathrm{L}}}P_{\mathrm{T}}\gamma_{\mathrm{T}}G_{\mathrm{T},x_{i}}e^{-k_a(f_{\mathrm{T}})d_{\mathrm{L},x_{i}}}d_{\mathrm{L},x_{i}}^{-\alpha_{\mathrm{L}}}\chi_{\mathrm{L},x_{i}}.
\label{eq:IN}
\end{equation}
\section{Association probabilities and Serving Distance Distributions}
\label{sec:association_serving}
Referring to the considered association policy, the UE can be served by either a LOS THz AP, a NLOS THz AP or an RF AP, respectively. To account for the three association events, we divide the sample space into three different events, $C_Q$, $Q=\{\mathrm{L},\mathrm{N},\mathrm{R}\}$, representing the events that the UE choose to associate with a LOS THz AP, a NLOS THz AP or an RF AP, respectively. To this extent, the association probability is defined as the probability of occurrence of the disjoint event $C_Q$, $Q=\{\mathrm{L},\mathrm{N},\mathrm{R}\}$ and we denote it as $A_{Q}$. We start this section by presenting first relevant distance distributions and exclusion regions expressions. These expressions are helpful in obtaining the association probabilities and the Laplace transforms of the interference powers. Next, we derive the association probabilities and the corresponding serving distance distributions for the three different association events.
\vspace{-0.5cm}
\subsection{Relevant Distance Distributions and Exclusion Regions}
As the APs are distributed according to a BPP in a finite disk $\mathcal{A}$ of radius $r_d$, a useful distance distribution is that from the reference UE to an arbitrary AP located at $\textbf{x}_{i}$. Thus, the probability density function (PDF) of the distance between the AP $\textbf{x}_{i}$ and the reference UE at $\textbf{v}_{0}=(v_0,0,0)$ is given by~\cite[eq. (7)]{Chetlur}
\begin{equation}
f_{\mathrm{Z}}(z)=\begin{cases}
f_{\mathrm{Z}_{1}}(z)=\frac{2z}{r_{d}^2}, & z_{l}\leq z \leq z_{m}\\
f_{\mathrm{Z}_{2}}(z)=\frac{2z}{\pi r_{d}^2}\arccos\left(\frac{z^2+v_{0}^2-r_d^2-(h_{\mathrm{A}}-h_{\mathrm{U}})^{2}}{2v_0 \sqrt{z^2-(h_{\mathrm{A}}-h_{\mathrm{U}})^2}}\right), & z_{m}\leq z \leq z_{p}
\end{cases},
\label{eq:fw}
\end{equation}
where $z_{l}=h_{\mathrm{A}}-h_{\mathrm{U}}$, $z_{m}=\sqrt{(r_d-v_0)^2+(h_{\mathrm{A}}-h_{\mathrm{U}})^2}$ $z_{p}=\sqrt{(r_d+v_0)^2+(h_{\mathrm{A}}-h_{\mathrm{U}})^2}$, and $h_{\mathrm{A}}$ and $h_{\mathrm{U}}$ are the heights at which the APs and the reference UE are deployed.

We can notice from the considered system model that all the APs are located at distances larger than $h_{\mathrm{A}}-h_{\mathrm{U}}$ from the reference UE. Furthermore, the association policy creates an exclusion region on the positions of the closest interfering AP and therefore on the positions of the remaining APs in each set and for each association event, i.e, for each event $C_Q$, $Q=\{\mathrm{L},\mathrm{N},\mathrm{R}\}$. The following remarks give the exclusion regions on the locations of the APs for each association type, respectively.
\begin{remark}
As the UE associates with a LOS THz AP located at a distance $r$, the nearest RF AP and NLOS AP are located further than $E_{\mathrm{LR}}(r)$ and $E_{\mathrm{LN}}(r)$, respectively, given by
\begin{equation}\small
E_{\mathrm{LR}}(r)=\begin{cases}h_{\mathrm{A}}-h_{\mathrm{U}}, & h_{\mathrm{A}}-h_{\mathrm{U}}\leq r <h_{\mathrm{LR}}\\
\left(\frac{P_{\mathrm{R}}\gamma_{\mathrm{R}}}{B_{\mathrm{T}} P_{\mathrm{T}} \gamma_{\mathrm{T}}G_{\mathrm{T},0}^{(\mathrm{mean})}}\right)^{\frac{1}{\alpha_{\mathrm{R}}}}e^{\frac{k_a(f)}{\alpha_{\mathrm{R}}}r}r^{\frac{\alpha_{\mathrm{L}}}{\alpha_{\mathrm{R}}}}, & r \geq h_{\mathrm{LR}},
\end{cases}
\label{eq:ELR}
\end{equation}
\begin{equation}\small
E_{\mathrm{LN}}(r)=\begin{cases}h_{\mathrm{A}}-h_{\mathrm{U}}, & h_{\mathrm{A}}-h_{\mathrm{U}}\leq r <h_{\mathrm{LN}}\\
\frac{\alpha_{\mathrm{N}}}{k_a(f)}\mathcal{W}\left[\frac{k_a(f)}{\alpha_{\mathrm{N}}}e^{\frac{k_a(f)}{\alpha_{\mathrm{N}}}r}r^{\frac{\alpha_{\mathrm{L}}}{\alpha_{\mathrm{N}}}}\right]& r \geq h_{\mathrm{LN}},
\end{cases}
\label{eq:ELN}
\end{equation}
where $\mathcal{W}[\cdot]$ is the Lambert W-function, $ h_{\mathrm{LR}}=\frac{\alpha_{\mathrm{L}}}{k_a(f)}\mathcal{W}\left[\frac{k_a(f)}{\alpha_{\mathrm{L}}}\left(\frac{B_{\mathrm{T}} P_{\mathrm{T}} \gamma_{\mathrm{T}}G_{\mathrm{T},0}^{\left(\mathrm{mean}\right)}}{P_{\mathrm{R}} \gamma_{\mathrm{R}}}\right)^{\frac{1}{\alpha_{\mathrm{L}}}}(h_{\mathrm{A}}-h_{\mathrm{U}})^{\frac{\alpha_{\mathrm{R}}}{\alpha_{\mathrm{L}}}}\right]$ and $h_{\mathrm{LN}}=\frac{\alpha_{\mathrm{L}}}{k_a(f)}\mathcal{W}\left[\frac{k_a(f)}{\alpha_{\mathrm{L}}}e^{\frac{k_a(f)}{\alpha_{\mathrm{N}}}(h_{\mathrm{A}}-h_{\mathrm{U}})}(h_{\mathrm{A}}-h_{\mathrm{U}})^{\frac{\alpha_{\mathrm{N}}}{\alpha_{\mathrm{L}}}}\right]$.
\begin{IEEEproof}
\emph{The biased average received power from a LOS UAV located at $r$ is given as $P_{\mathrm{L}}^{r}=B_{\mathrm{T}}P_{\mathrm{T}}\gamma_{\mathrm{T}}G_{\mathrm{T},0}^{\left(\mathrm{mean}\right)}e^{-k_{a}(f)r}r^{-\alpha_{\mathrm{L}}}$. As the UE associates to the AP that provides the strongest BRSP, the distance separating the closest RF AP from the UE $E_{\mathrm{LR}}(r)$ is obtained by solving the equation $B_{\mathrm{T}}P_{\mathrm{T}}\gamma_{\mathrm{T}}G_{\mathrm{T},0}^{\left(\mathrm{mean}\right)}e^{-k_{a}(f)r}r^{-\alpha_{\mathrm{L}}}=P_{\mathrm{R}}\gamma_{\mathrm{R}}E_{\mathrm{LR}}^{-\alpha_{\mathrm{R}}}(r)$. Similarly, $E_{\mathrm{LN}}(r)$ can be found by solving the equation $B_{\mathrm{T}}P_{\mathrm{T}}\gamma_{\mathrm{T}}G_{\mathrm{T},0}^{\left(\mathrm{mean}\right)}e^{-k_{a}(f)r}r^{-\alpha_{\mathrm{L}}}=B_{\mathrm{T}}P_{\mathrm{T}}\gamma_{\mathrm{T}}G_{\mathrm{T},0}^{\left(\mathrm{mean}\right)}e^{-k_{a}(f)E_{\mathrm{LN}}(r)}E_{\mathrm{LN}}^{-\alpha_{\mathrm{L}}}(r)$. The above interpretation holds if $h_{\mathrm{LR}}>h_{\mathrm{A}}-h_{\mathrm{U}}$ and $h_{\mathrm{LN}}>h_{\mathrm{A}}-h_{\mathrm{U}}$. Otherwise, the expressions of $E_{\mathrm{LR}}(r)$ and $E_{\mathrm{LN}}(r)$ are simplified to the second term of the piece-wise functions for $r\geq h_{\mathrm{A}}-h_{\mathrm{U}}$.}
\end{IEEEproof}
\end{remark}
\vspace{-1cm}
\begin{remark}
As the UE associates with a NLOS AP located at a distance $r$, the nearest RF and LOS APs are located further than $E_{\mathrm{NL}}(r)$ and $E_{\mathrm{NR}}(r)$, respectively, given by
\begin{equation}\small
E_{\mathrm{NR}}(r)=\begin{cases}h_{\mathrm{A}}-h_{\mathrm{U}}, & h_{\mathrm{A}}-h_{\mathrm{U}}\leq r <h_{\mathrm{NR}}\\
\left(\frac{P_{\mathrm{R}}\gamma_{\mathrm{R}}}{B_{\mathrm{T}} P_{\mathrm{T}} \gamma_{\mathrm{T}}G_{\mathrm{T},0}^{(\mathrm{mean})}}\right)^{\frac{1}{\alpha_{\mathrm{R}}}}e^{\frac{k_a(f)}{\alpha_{\mathrm{R}}}r}r^{\frac{\alpha_{\mathrm{N}}}{\alpha_{\mathrm{R}}}}, & r \geq h_{\mathrm{NR}},
\end{cases}
\label{eq:ENR}
\end{equation}
\begin{equation}\small
E_{\mathrm{NL}}(r)=\begin{cases}h_{\mathrm{A}}-h_{\mathrm{U}}, & h_{\mathrm{A}}-h_{\mathrm{U}}\leq r <h_{\mathrm{NL}}\\
\frac{\alpha_{\mathrm{L}}}{k_a(f)}\mathcal{W}\left[\frac{k_a(f)}{\alpha_{\mathrm{L}}}e^{\frac{k_a(f)}{\alpha_{\mathrm{L}}}r}r^{\frac{\alpha_{\mathrm{N}}}{\alpha_{\mathrm{L}}}}\right]& r \geq h_{\mathrm{NL}},
\end{cases}
\label{eq:ENL}
\end{equation}
where $\mathcal{W}[\cdot]$ is the Lambert W-function, $h_{\mathrm{NR}}=\frac{\alpha_{\mathrm{N}}}{k_a(f)}\mathcal{W}\left[\frac{k_a(f)}{\alpha_{\mathrm{N}}}\left(\frac{B_{\mathrm{T}} P_{\mathrm{T}} \gamma_{\mathrm{T}}G_{\mathrm{T},0}^{\left(\mathrm{mean}\right)}}{P_{\mathrm{R}} \gamma_{\mathrm{R}}}\right)^{\frac{1}{\alpha_{\mathrm{N}}}}(h_{\mathrm{A}}-h_{\mathrm{U}})^{\frac{\alpha_{\mathrm{R}}}{\alpha_{\mathrm{N}}}}\right]$ and $h_{\mathrm{NL}}=\frac{\alpha_{\mathrm{N}}}{k_a(f)}\mathcal{W}\left[\frac{k_a(f)}{\alpha_{\mathrm{N}}}e^{\frac{k_a(f)}{\alpha_{\mathrm{L}}}(h_{\mathrm{A}}-h_{\mathrm{U}})}(h_{\mathrm{A}}-h_{\mathrm{U}})^{\frac{\alpha_{\mathrm{L}}}{\alpha_{\mathrm{N}}}}\right]$. 
\begin{IEEEproof}
\emph{The proof follows the same steps of obtaining (\ref{eq:ELR}) and (\ref{eq:ELN}), therefore omitted here.}
\end{IEEEproof}
\end{remark}
\vspace{-0.7cm}
\begin{remark}
	 As the UE associates to an RF AP located at a distance $r$, the nearest LOS and NLOS APs are located further than $E_{\mathrm{RL}}(r)$ and $E_{\mathrm{RN}}(r)$, respectively, given by
	\begin{equation}\small
	E_{\mathrm{RL}}(r)=\begin{cases}h_{\mathrm{A}}-h_{\mathrm{U}}, & h_{\mathrm{A}}-h_{\mathrm{U}}\leq r <h_{\mathrm{RL}}\\
	\frac{\alpha_{\mathrm{L}}}{k_a(f)}\mathcal{W}\left[\frac{k_a(f)}{\alpha_{\mathrm{L}}}\left(\frac{B_{\mathrm{T}} P_{\mathrm{T}} \gamma_{\mathrm{T}}G_{\mathrm{T},0}^{\left(\mathrm{mean}\right)}}{P_{\mathrm{R}} \gamma_{\mathrm{R}}}\right)^{\frac{1}{\alpha_{\mathrm{L}}}}r^{\frac{\alpha_{\mathrm{R}}}{\alpha_{\mathrm{L}}}}\right], & r \geq h_{\mathrm{RL}},
	\end{cases}
	\label{eq:ERL}
	\end{equation}
	\begin{equation}\small
	E_{\mathrm{RN}}(r)=\begin{cases}h_{\mathrm{A}}-h_{\mathrm{U}}, & h_{\mathrm{A}}-h_{\mathrm{U}}\leq r <h_{\mathrm{RN}}\\
	\frac{\alpha_{\mathrm{N}}}{k_a(f)}\mathcal{W}\left[\frac{k_a(f)}{\alpha_{\mathrm{N}}}\left(\frac{B_{\mathrm{T}} P_{\mathrm{T}} \gamma_{\mathrm{T}}G_{\mathrm{T},0}^{\left(\mathrm{mean}\right)}}{P_{\mathrm{R}} \gamma_{\mathrm{R}}}\right)^{\frac{1}{\alpha_{\mathrm{N}}}}r^{\frac{\alpha_{\mathrm{R}}}{\alpha_{\mathrm{N}}}}\right], & r \geq h_{\mathrm{RN}},
	\end{cases}
	\label{eq:ERN}
	\end{equation}
	where $\mathcal{W}[\cdot]$ is the Lambert W-function, $h_{\mathrm{RL}}=\left(\frac{P_{\mathrm{R}}\gamma_{\mathrm{R}}}{B_{\mathrm{T}} P_{\mathrm{T}} \gamma_{\mathrm{T}}G_{\mathrm{T},0}^{(\mathrm{mean})}}\right)^{\frac{1}{\alpha_{\mathrm{R}}}}e^{\frac{k_a(f)}{\alpha_{\mathrm{R}}}(h_{\mathrm{A}}-h_{\mathrm{U}})}(h_{\mathrm{A}}-h_{\mathrm{U}})^{\frac{\alpha_{\mathrm{L}}}{\alpha_{\mathrm{R}}}}$ and $h_{\mathrm{RN}}=\left(\frac{P_{\mathrm{R}}\gamma_{\mathrm{R}}}{B_{\mathrm{T}} P_{\mathrm{T}} \gamma_{\mathrm{T}}G_{\mathrm{T},0}^{(\mathrm{mean})}}\right)^{\frac{1}{\alpha_{\mathrm{R}}}}e^{\frac{k_a(f)}{\alpha_{\mathrm{R}}}(h_{\mathrm{A}}-h_{\mathrm{U}})}(h_{\mathrm{A}}-h_{\mathrm{U}})^{\frac{\alpha_{\mathrm{N}}}{\alpha_{\mathrm{R}}}}$.
	\begin{IEEEproof}
		\emph{The proof is similar to that of (\ref{eq:ELR}) and (\ref{eq:ELN}), therefore omitted here.}
	\end{IEEEproof}
\end{remark}
\vspace{-0.5cm}
\subsection{Association Probabilities}
In the coexisting RF/THz network, the UE associates with an RF AP, a LOS THz AP or a NLOS THz AP according to the maximum BRSP association policy. The corresponding association probabilities $A_{\mathrm{L}}$, $A_{\mathrm{N}}$ and $A_{\mathrm{R}}$ are presented in the subsequent lemmas.
\begin{lemma}
	The probability that the reference UE is served by a LOS THz AP, denoted as the LOS THz association probability $A_{\mathrm{L}}$, is calculated as
	\begin{equation}\small
	\begin{aligned}
	&A_{\mathrm{L}}=\delta_{\mathrm{T}}N_{\mathrm{A}}\\
	&\int_{z_{l}}^{z_{p}}f_{\mathrm{Z}}(r)\kappa_{\mathrm{L}}(r)\left(\int_{E_{\mathrm{LR}}(r)}^{z_{p}}f_{\mathrm{Z}}(z)\mathrm{d}z\right)^{(1-\delta_{\mathrm{T}})N_{\mathrm{A}}}\left(\int_{r}^{z_{p}}f_{\mathrm{Z}}(z)\kappa_{\mathrm{L}}(z)\mathrm{d}z+\int_{E_{\mathrm{LN}}(r)}^{z_{p}}f_{\mathrm{Z}}(z)\kappa_{\mathrm{N}}(z)\mathrm{d}z\right)^{\delta_{\mathrm{T}} N_{\mathrm{A}}-1}\mathrm{d}r,\\
	\end{aligned}
	\label{eq:AL}
	\end{equation}
	where $N_{\mathrm{A}}$ is the number of APs, $\delta_{\mathrm{T}}$ is the percentage of THz APs, $f_{\mathrm{Z}}(z)$ is the PDF of distance from the UE to any AP given in (\ref{eq:fw}), $\kappa_{\mathrm{L}}(\cdot)$ and $\kappa_{\mathrm{N}}(\cdot)$ are the LOS and NLOS probabilities given in (\ref{eq:PL}) and (\ref{eq:PN}), respectively. $E_{\mathrm{LR}}(r)$, $E_{\mathrm{LN}}(r)$ represent the exclusion regions on the locations of the RF and the NLOS THz APs and can be found in (\ref{eq:ELR}) and (\ref{eq:ELN}), respectively.
 \begin{IEEEproof}
	\emph{See Appendix~\ref{app:AL}.}
\end{IEEEproof} 
\label{lemma:AL}
\end{lemma}
\begin{lemma}
		The probability that the reference UE is served by a NLOS THz AP, denoted as the NLOS THz association probability $A_{\mathrm{N}}$, is calculated as
	\begin{equation}\small
	\begin{aligned}
	&A_{\mathrm{N}}=\delta_{\mathrm{T}} N_{\mathrm{A}}\\
	&\int_{z_{l}}^{z_{p}}f_{\mathrm{Z}}(r)\kappa_{\mathrm{N}}(r)\left(\int_{E_{\mathrm{NR}}(r)}^{z_{p}}f_{\mathrm{Z}}(z)\mathrm{d}z\right)^{(1-\delta_T)N_{\mathrm{A}}}\left(\int_{r}^{z_{p}}f_{\mathrm{Z}}(z)\kappa_{\mathrm{N}}(z)\mathrm{d}z+\int_{E_{\mathrm{NL}}(r)}^{z_{p}}f_{\mathrm{Z}}(z)\kappa_{\mathrm{L}}(z)\mathrm{d}z\right)^{\delta_{\mathrm{T}} N_{\mathrm{A}}-1}\mathrm{d}r,\\
	\end{aligned}
	\label{eq:AN}
	\end{equation}
	where $E_{\mathrm{NR}}(r)$, $E_{\mathrm{NL}}(r)$ are the exclusion regions on the locations of the RF APs and the LOS THz APs and are provided in (\ref{eq:ENR}) and (\ref{eq:ENL}), respectively.
	 \begin{IEEEproof}
	\emph{The proof refers to a similar procedure as in Lemma~\ref{lemma:AL}, therefore omitted here.}
	\end{IEEEproof} 
	\label{lemma:AN}
\end{lemma}
\begin{lemma}
	The probability that the reference UE is served by an RF AP, denoted as the RF association probability $A_{\mathrm{R}}$, is given by
	\begin{equation}\small
	\begin{aligned}
	&A_{\mathrm{R}}=(1-\delta_{\mathrm{T}}) N_{\mathrm{A}}\\
	&\int_{z_{l}}^{z_{p}}f_{\mathrm{Z}}(r)\left(\int_{r}^{z_{p}}f_{\mathrm{Z}}(z)\mathrm{d}z\right)^{(1-\delta_T)N_{\mathrm{A}}-1}\left(\int_{E_{\mathrm{RL}}(r)}^{z_{p}}f_{\mathrm{Z}}(z)\kappa_{\mathrm{L}}(z)\mathrm{d}z+\int_{E_{\mathrm{RN}}(r)}^{z_{p}}f_{\mathrm{Z}}(z)\kappa_{\mathrm{N}}(z)\mathrm{d}z\right)^{\delta_T N_{\mathrm{A}}}\mathrm{d}r,\\
	\end{aligned}
	\label{eq:AR}
	\end{equation}
    where $E_{\mathrm{RL}}(r)$, and $E_{\mathrm{RN}}(r)$ represent the exclusion regions on the positions of the LOS THz APs and the NLOS THz APs and are provided in~(\ref{eq:ERL}) and~(\ref{eq:ERN}), respectively.
  	 \begin{IEEEproof}
  	\emph{The proof of this lemma refers to a similar procedure as in Lemma~\ref{lemma:AL}, therefore omitted here.}
  \end{IEEEproof} 
  \label{lemma:AR}
\end{lemma}
 Note here that the probability of the UE being associated with a THz AP is given by $A_{\mathrm{T}}=A_{\mathrm{L}}+A_{\mathrm{N}}$ and that $A_{\mathrm{T}}+A_{\mathrm{R}}=1$.
 \vspace{-0.5cm}
\subsection{Serving Distance Distributions}
In this section, we present the conditional distance distributions separating the UE from the serving LOS THz AP, NLOS THz AP and RF AP. The derived distance distributions are provided in lemmas $4$, $5$ and $6$, respectively.
\begin{lemma}
	The PDF of the distance separating the location of the UE from its serving AP, being a LOS THz AP, denoted by $f_{X_{\mathrm{L}}}(\cdot)$ can be obtained as
\begin{equation}\small
\begin{aligned}
&f_{X_{\mathrm{L}}}(x_{\mathrm{L}})=\frac{\delta_{\mathrm{T}} N_{\mathrm{A}}}{A_{\mathrm{L}}}f_{\mathrm{Z}}(x_{\mathrm{L}})\kappa_{\mathrm{L}}(x_{\mathrm{L}})\\
&\times\left(\int_{E_{\mathrm{LR}}(x_{\mathrm{L}})}^{z_{p}}f_{\mathrm{Z}}(z)\mathrm{d}z\right)^{(1-\delta_{\mathrm{T}})N_{\mathrm{A}}}\left(\int_{x_{\mathrm{L}}}^{z_{p}}f_{\mathrm{Z}}(z)\kappa_{\mathrm{L}}(z)\mathrm{d}z+\int_{E_{\mathrm{LN}}(x_{\mathrm{L}})}^{z_{p}}f_{\mathrm{Z}}(z)\kappa_{\mathrm{N}}(z)\mathrm{d}z\right)^{\delta_{\mathrm{T}} N_{\mathrm{A}}-1},\\
\end{aligned}
\label{eq:fxL}
\end{equation}
where $f_{\mathrm{Z}}(\cdot)$, $E_{\mathrm{LR}}(\cdot)$, $E_{\mathrm{LN}}(\cdot)$ and $A_{\mathrm{L}}$ are provided in (\ref{eq:fw}), (\ref{eq:ELR}), (\ref{eq:ELN}), and (\ref{eq:AL}), respectively.
\begin{IEEEproof}
	See Appendix~\ref{app:servingdistance}.
\end{IEEEproof}
\label{lemma:xL}
\end{lemma}
\begin{lemma}
    The PDF of the distance separating the location of the UE from its serving AP, being a NLOS THz AP, denoted by $f_{X_{\mathrm{N}}}(\cdot)$ can be obtained as
	\begin{equation}\small
	\begin{aligned}
	&f_{X_{\mathrm{N}}}(x_{\mathrm{N}})=\frac{\delta_{\mathrm{T}} N_{\mathrm{A}}}{A_{\mathrm{N}}}f_{\mathrm{Z}}(x_{\mathrm{N}})\kappa_{\mathrm{N}}(x_{\mathrm{N}})\\
	&\times\left(\int_{E_{\mathrm{NR}}(x_{\mathrm{N}})}^{z_{p}}f_{\mathrm{Z}}(z)\mathrm{d}z\right)^{(1-\delta_{\mathrm{T}})N_{\mathrm{A}}}\left(\int_{x_{\mathrm{N}}}^{z_{p}}f_{\mathrm{Z}}(z)\kappa_{\mathrm{N}}(z)\mathrm{d}z+\int_{E_{\mathrm{NL}}(x_{\mathrm{N}})}^{z_{p}}f_{\mathrm{Z}}(z)\kappa_{\mathrm{L}}(z)\mathrm{d}z\right)^{\delta_{\mathrm{T}} N_{\mathrm{A}}-1},\\
	\end{aligned}
	\label{eq:fxN}
	\end{equation}
	where $E_{\mathrm{NR}}(\cdot)$, $E_{\mathrm{NL}}(\cdot)$ and $A_{\mathrm{N}}$ are provided in (\ref{eq:ENR}), (\ref{eq:ENL}), and (\ref{eq:AN}), respectively.
	\begin{IEEEproof}
	\emph{This proof follows a similar procedure as in Lemma~\ref{lemma:xL}, therefore omitted here.}
	\end{IEEEproof}
	\label{lemma:xN}
\end{lemma}
\begin{lemma}
	The PDF of the distance separating the location of the UE from its serving AP, being an RF AP, denoted by $f_{X_{\mathrm{R}}}(\cdot)$ can be obtained as
	\begin{equation}\small
	\begin{aligned}
	&f_{X_{\mathrm{R}}}(x_{\mathrm{R}})=\frac{(1-\delta_{\mathrm{T}}) N_{\mathrm{A}}}{A_{\mathrm{R}}}f_{\mathrm{Z}}(x_{\mathrm{R}})\\
	&\left(\int_{x_{\mathrm{R}}}^{z_{p}}f_{\mathrm{Z}}(z)\mathrm{d}z\right)^{(1-\delta_{\mathrm{T}})N_{\mathrm{A}}-1}\left(\int_{E_{\mathrm{RL}}(x_{\mathrm{R}})}^{z_{p}}f_{\mathrm{Z}}(z)\kappa_{\mathrm{L}}(z)\mathrm{d}z+\int_{E_{\mathrm{RN}}(x_{\mathrm{R}})}^{z_{p}}f_{\mathrm{Z}}(z)\kappa_{\mathrm{N}}(z)\mathrm{d}z\right)^{\delta_{\mathrm{T}} N_{\mathrm{A}}},\\
	\end{aligned}
	\label{eq:fxR}
	\end{equation}
 where $E_{\mathrm{RL}}(\cdot)$, $E_{\mathrm{RN}}(\cdot)$ and $A_{\mathrm{R}}$ are provided in~(\ref{eq:ERL}), (\ref{eq:ERN}), and~(\ref{eq:AR}), respectively.
 	\begin{IEEEproof}
	\emph{This proof follows a similar procedure as in Lemma~\ref{lemma:xL}, therefore omitted here.}
	\end{IEEEproof}
	\label{lemma:xR}
\end{lemma}
\vspace{-0.7cm}
\section{Coverage Probability and Average Achievable Rate}
\label{sec:coverage_rate}
\subsection{Coverage probability}
The coverage probability is defined as the probability that the $\mathrm{SINR}$ at the UE exceeds a predefined threshold. Since a UE can associate with a LOS THz AP, a NLOS THz AP or an RF AP, the coverage probability can be calculated by referring to the law of total probability as
\begin{equation}
P_{cov}=A_{\mathrm{L}} P_{cov,\mathrm{L}}+A_{\mathrm{N}} P_{cov,\mathrm{N}}+A_{\mathrm{R}} P_{cov,\mathrm{R}},
\end{equation}
where $A_{\mathrm{L}}$, $A_{\mathrm{N}}$ and $A_{\mathrm{R}}$ are the corresponding association probabilities (i.e. the probability of occurrence of the event $C_{Q}$, $Q=\{\mathrm{L}, \mathrm{N}, \mathrm{R}\}$) given in (\ref{eq:AL}), (\ref{eq:AN}) and (\ref{eq:AR}) and $P_{cov,\mathrm{L}}$, $P_{cov,\mathrm{N}}$ and $P_{cov,\mathrm{R}}$ are the conditional coverage probabilities given the association status and are provided in the following theorem.
 \begin{theorem}
 The conditional coverage probabilities $P_{cov,\mathrm{L}}$, $P_{cov,\mathrm{N}}$ and $P_{cov,\mathrm{R}}$ given that the UE is associated with a LOS THz AP, a NLOS THz AP and RF AP are given by
 \begin{equation}\small
 P_{cov,\mathrm{L}}=\sum_{k=1}^{4}p_{k,0}\mathbb{E}_{x_{\mathrm{L}}}\left[\sum_{q=0}^{m_{\mathrm{L}}-1}\frac{(-s_{\mathrm{L}}(x_{\mathrm{L}}))^{q}}{q!}\sum_{u=0}^{q}{q\choose u}\left(-\frac{\sigma_{\mathrm{T}}^{2}}{G_{k}}\right)^{(q-u)}\exp\left(-\frac{s_{\mathrm{L}}(x_{\mathrm{L}})\sigma_{\mathrm{T}}^2}{G_{k}}\right)\frac{\partial^{u}}{\partial s_{\mathrm{L}}^{u}}\mathcal{L}_{I_{\mathrm{L}}}\left(\frac{s_{\mathrm{L}}(x_{\mathrm{L}})}{G_{k}}\right)\right],
 \label{eq:PcovL}
 \end{equation}
 \begin{equation}\small
 P_{cov,\mathrm{N}}
 =\sum_{k=1}^{4}p_{k,0}\mathbb{E}_{x_{\mathrm{N}}}\left[\sum_{q=0}^{m_{\mathrm{N}}-1}\frac{(-s_{\mathrm{N}}(x_{\mathrm{N}}))^{q}}{q!}\sum_{u=0}^{q}{q\choose u}\left(-\frac{\sigma_{\mathrm{T}}^{2}}{G_{k}}\right)^{(q-u)}\exp\left(-\frac{s_{\mathrm{N}}(x_{\mathrm{N}})\sigma_{\mathrm{T}}^2}{G_{k}}\right)\frac{\partial^{u}}{\partial s_{\mathrm{N}}^{u}}\mathcal{L}_{I_{\mathrm{N}}}\left(\frac{s_{\mathrm{N}}(x_{\mathrm{N}})}{G_{k}}\right)\right], 
 \label{eq:PcovN}
 \end{equation}
  \vspace{-0.5cm}
 \begin{equation}
 P_{cov,\mathrm{R}}=\mathbb{E}_{x_{\mathrm{R}}}\left[\mathcal{L}_{I_{\mathrm{R}}}\left(s_{\mathrm{R}}(x_{\mathrm{R}})\right)\exp\left(-s_{\mathrm{R}}\left(x_{\mathrm{R}}\right) \sigma_{\mathrm{R}}^2\right)\right],
 \label{eq:PcovR}
 \end{equation}
 where $s_{\mathrm{R}}(x_{\mathrm{R}})=\frac{\theta}{P_{\mathrm{R}}\gamma_{\mathrm{R}}x_{\mathrm{R}}^{-\alpha_{\mathrm{R}}}}$, $s_{\mathrm{L}}(x_{\mathrm{L}})=\frac{m_{\mathrm{L}}\theta e^{k_{a}(f_{\mathrm{T}})x_{\mathrm{L}}}x_{\mathrm{L}}^{\alpha_{\mathrm{L}}}}{P_{\mathrm{T}}\gamma_{\mathrm{T}}}$ and $s_{\mathrm{N}}(x_{\mathrm{N}})=\frac{m_{\mathrm{N}}\theta e^{k_{a}(f_{\mathrm{T}})x_{\mathrm{N}}}x_{\mathrm{N}}^{\alpha_{\mathrm{N}}}}{P_{\mathrm{T}}\gamma_{\mathrm{T}}}$. $\mathcal{L}_{I_{\mathrm{L}}}(\cdot)$, $\mathcal{L}_{I_{\mathrm{N}}}(\cdot)$ and $\mathcal{L}_{I_{\mathrm{R}}}(\cdot)$ represent the Laplace transforms of the interference in the three considered association scenarios.
 \begin{IEEEproof}
 See Appendix~\ref{app:coverage}.
 \end{IEEEproof}
	\label{theorem:coverage}
	\vspace{-0.3cm}
 \end{theorem}
To be able to derive the expressions of the conditional coverage probabilities, we must find the Laplace transforms of the interference in the three considered association scenarios. Due to the separate spectrum for THz and RF, the UE receives interference signals from the THz APs only when associated with a LOS or a NLOS THz AP. Similarly, the UE receives interference from the RF APs only when associated with an RF AP. The interference expressions $I_{\mathrm{L}}$, $I_{\mathrm{N}}$ and $I_{\mathrm{R}}$ are given in (\ref{eq:IL}), (\ref{eq:IN}) and (\ref{eq:IR}), respectively. The Laplace transforms of $I_{\mathrm{L}}$, $I_{\mathrm{N}}$ and $I_{\mathrm{R}}$ are given in the following lemmas. 
\vspace{-0.3cm}
\begin{lemma}
The Laplace transform of the interference power $I_{\mathrm{L}}$ from the LOS and NLOS THz APs given that the UE is associated with a LOS THz AP can be given as
\begin{equation}\small
\begin{aligned}
&\mathcal{L}_{I_{\mathrm{L}}}(s)=\\&\left[\frac{1}{\int_{x_{\mathrm{L}}}^{z_{p}}f_{\mathrm{Z}}(z)\kappa_{\mathrm{L}}(z)\mathrm{d}z+\int_{E_{\mathrm{LN}}\left(x_{\mathrm{L}}\right)}^{z_{p}}f_{\mathrm{Z}}(z)\kappa_{\mathrm{N}}(z)\mathrm{d}z}\right.
\sum_{k=1}^{4}p_{k}\left(\int_{x_{\mathrm{L}}}^{z_{p}}f_{\mathrm{Z}}(y)\kappa_{\mathrm{L}}(y)\left(1+\frac{s P_{\mathrm{T}}\gamma_{\mathrm{T}}G_{k}e^{-k_a(f_{\mathrm{T}})y}y^{-\alpha_{\mathrm{L}}}}{m_{\mathrm{L}}}\right)^{-m_{\mathrm{L}}}\mathrm{d}y\right.\\
&\left.\left.+\int_{E_{\mathrm{LN}}\left(x_{\mathrm{L}}\right)}^{z_{p}}f_{\mathrm{Z}}(y)\kappa_{\mathrm{N}}(y)\left(1+\frac{s P_{\mathrm{T}}\gamma_{\mathrm{T}}G_{k}e^{-k_a(f)y}y^{-\alpha_{\mathrm{N}}}}{m_{\mathrm{N}}}\right)^{-m_{\mathrm{N}}}\mathrm{d}y\right)\right]^{\delta_{\mathrm{T}} N_{\mathrm{A}}-1}.
\end{aligned}
\label{eq:LIL}
\vspace{-0.3cm}
\end{equation}
\begin{IEEEproof}
	\emph{See Appendix~\ref{app:LIL}.}
\end{IEEEproof}
\label{lemma:LIL}
\end{lemma}
\vspace{-0.5cm}
\begin{lemma}
	 The Laplace transform of the interference power $I_{\mathrm{N}}$ from the NLOS and LOS THz APs given that the UE is associated with a NLOS THz AP can be given as
\begin{equation}\small
\begin{aligned}
&\mathcal{L}_{I_{\mathrm{N}}}(s)=\\&\left[\frac{1}{\int_{x_{\mathrm{N}}}^{z_{p}}f_{\mathrm{Z}}(z)\kappa_{\mathrm{N}}(z)\mathrm{d}z+\int_{E_{\mathrm{NL}}\left(x_{\mathrm{N}}\right)}^{z_{p}}f_{\mathrm{Z}}(z)\kappa_{\mathrm{L}}(z)\mathrm{d}z}\right.
\sum_{k=1}^{4}p_{k}\left(\int_{x_{\mathrm{N}}}^{z_{p}}f_{\mathrm{Z}}(y)\kappa_{\mathrm{N}}(y)\left(1+\frac{s P_{\mathrm{T}}\gamma_{\mathrm{T}}G_{k}e^{-k_a(f_{\mathrm{T}})y}y^{-\alpha_{\mathrm{N}}}}{m_{\mathrm{N}}}\right)^{-m_{\mathrm{N}}}\mathrm{d}y\right.\\
&\left.\left.+\int_{E_{\mathrm{NL}}\left(x_{\mathrm{N}}\right)}^{z_{p}}f_{\mathrm{Z}}(y)\kappa_{\mathrm{L}}(y)\left(1+\frac{s P_{\mathrm{T}}\gamma_{\mathrm{T}}G_{k}e^{-k_a(f)y}y^{-\alpha_{\mathrm{L}}}}{m_{\mathrm{L}}}\right)^{-m_{\mathrm{L}}}\mathrm{d}y\right)\right]^{\delta_{\mathrm{T}} N_{\mathrm{A}}-1}.
\end{aligned}
\label{eq:LIN}
\vspace{-0.3cm}
\end{equation}
\begin{IEEEproof}
	\emph{This proof follows the same procedure used in Lemma~\ref{lemma:LIL}, therefore omitted here.}
\end{IEEEproof}
\label{lemma:LIN}
\end{lemma}
\begin{lemma}
	Given that the UE associates with an RF AP, the Laplace transform of the interference power $I_{\mathrm{R}}$ from the remaining RF APs is given by
\begin{equation}
\mathcal{L}_{I_{\mathrm{R}}}(s)=\left(\frac{1}{\int_{x_\mathrm{R}}^{z_{p}}f_{\mathrm{Z}}(z)\mathrm{d}z}\int_{x_\mathrm{R}}^{z_{p}}\frac{1}{1+s P_{\mathrm{R}} \gamma_{\mathrm{R}}y^{-\alpha_\mathrm{R}}}f_{\mathrm{Z}}(y)\mathrm{d}y\right)^{\left(1-\delta_{\mathrm{T}}\right)N_{\mathrm{A}}-1}.
\label{eq:LIR}
\end{equation}
\begin{IEEEproof}
\emph{The interference $I_{\mathrm{R}}$ from all interfering RF APs when the UE associates with an RF AP located at $x_{\mathrm{R}}$ can be written as $\sum_{i=1}^{(1-\delta_{\mathrm{T}})N_{\mathrm{A}}-1}I_{\mathrm{R},x_{i}}$, where $I_{\mathrm{R},x_{i}}$ is the interference from the RF AP located at $\textbf{x}_{i}$. Following a similar proof to lemma~\ref{lemma:LIL}, the Laplace transform of $I_{\mathrm{R}}$, denoted as $\mathcal{L}_{I_{\mathrm{R}}}(s)$ can be expressed as
\begin{equation}
\mathcal{L}_{I_{\mathrm{R}}}(s)=\left(\mathbb{E}_{I_{\mathrm{R},x_{i}}}\left[\exp\left(-s I_{\mathrm{R},x_{i}}\right)\right]\right)^{(\delta_{T}-1)N_{\mathrm{A}}-1}. \label{eq:LIR_proof1}   
\end{equation}
The expectation term can be calculated as
\begin{equation}
\begin{aligned}
     \mathbb{E}_{I_{\mathrm{R},x_{i}}}\left[\exp\left(-s I_{\mathrm{R},x_{i}}\right)\right]\stackrel{(a)}{=}\mathbb{E}_{\chi_{\mathrm{R}},d_{\mathrm{R}}}\left[\exp\left(-s P_{\mathrm{R}}\gamma_{\mathrm{R}}d_{\mathrm{R}}^{-\alpha_{\mathrm{R}}}\chi_{\mathrm{R}}\right)\right]
     \stackrel{(b)}{=}\int_{x_{\mathrm{R}}}^{z_{p}}\frac{1}{1+s P_{\mathrm{R}}\gamma_{\mathrm{R}}y_{-\alpha_{\mathrm{R}}}}f_{Y_{\mathrm{R}}}(y,x_{\mathrm{R}})\mathrm{d}y.
\end{aligned}
\label{eq:LIR_proof2}
\end{equation}
where (a) is obtained by omitting the index $x_{i}$ from the expression of the received power from an RF AP given in Section~\ref{sec:RF_channel_model}. (b) is obtained from the moment generating function (MGF) of the Rayleigh small scale fading gain and by substituting $d_{\mathrm{R}}$ with $y$ and averaging over the feasibility range of $y$. $f_{Y_{\mathrm{R}}}(y,x_{\mathrm{R}})$ is the PDF of the distance from any interfering RF AP located further than $x_{\mathrm{R}}$ and is given as $f_{Y_{\mathrm{R}}}(y,x)=\frac{f_{\mathrm{Z}}(y)}{\int_{x}^{z_{p}}f_{\mathrm{Z}}(z)\mathrm{d}z}$, where $f_{\mathrm{Z}}(.)$ is given in (\ref{eq:fw})~\cite{Chetlur}. By plugging (\ref{eq:LIR_proof2}) in (\ref{eq:LIR_proof1}), we can get the final expression in (\ref{eq:LIR}),
}
\end{IEEEproof}
\label{lemma:LIR}
\end{lemma}
\vspace{-0.8cm}
\subsection{Rate Analysis}
In this section, we derive the average achievable rate by using the same analysis conducted for the coverage probability. Thus, the average achievable rate is given in Theorem~\ref{theorem:rate}.
\vspace{-0.2cm}
\begin{theorem}
The average achievable DL rate of a UE located at a distance $v_0$ from the center of a THz and RF coexisting finite wireless network is given by
\begin{equation}
    \tau=\tau_{\mathrm{L}}A_{\mathrm{L}}+\tau_{\mathrm{N}}A_{\mathrm{N}}+\tau_{\mathrm{R}}A_{\mathrm{R}},
\end{equation}
where $A_{\mathrm{L}}$, $A_{\mathrm{N}}$, and $A_{\mathrm{R}}$ are the association probabilities and $\tau_{\mathrm{L}}$, $\tau_{\mathrm{N}}$, and $\tau_{\mathrm{R}}$ are the average achievable rates given that the UE associates with a LOS THz AP, NLOS THz AP, or RF AP, respectively, and are given by
\begin{equation}\small
\begin{aligned}
   \tau_{\mathrm{L}}&=\frac{W_{\mathrm{T}}}{\ln{2}}\sum_{k=1}^{4} p_{k,0}\\
   &\int_{0}^{\infty}\frac{1}{t+1}\mathbb{E}_{x_{\mathrm{L}}}\bigg[\sum_{q=0}^{m_{\mathrm{L}}-1}\frac{(-s_{\mathrm{L}}(x_{\mathrm{L}}))^{q}}{q!}\sum_{u=0}^{q}{q\choose u}\left(-\frac{\sigma_{\mathrm{T}}^{2}t}{G_{k}\theta}\right)^{q-u}\exp\left(-\frac{s_{\mathrm{L}}(x_{\mathrm{L}})\sigma_{\mathrm{T}}^2 t}{G_{k}\theta}\right)\frac{\partial^{u}}{\partial s_{\mathrm{L}}^{u}}\mathcal{L}_{I_{\mathrm{L}}}\left(\frac{s_{\mathrm{L}}(x_{\mathrm{L}})t}{G_{k}\theta}\right)\bigg]\mathrm{d}t,
\end{aligned}
\label{eq:tauL}
\vspace{-1cm}
\end{equation}
\begin{equation}\small
\begin{aligned}
\tau_{\mathrm{N}}&=\frac{W_{\mathrm{T}}}{\ln{2}}\sum_{k=1}^{4} p_{k,0}\\
&\int_{0}^{\infty}\frac{1}{t+1}\mathbb{E}_{x_{\mathrm{N}}}\bigg[\sum_{q=0}^{m_{\mathrm{N}}-1}\frac{(-s_{\mathrm{N}}(x_{\mathrm{N}}))^{q}}{q!}\sum_{u=0}^{q}{q\choose u}\left(-\frac{\sigma_{\mathrm{T}}^{2}t}{G_{k}\theta}\right)^{q-u}\exp\left(-\frac{s_{\mathrm{N}}(x_{\mathrm{N}})\sigma_{\mathrm{T}}^2 t}{G_{k}\theta}\right)\frac{\partial^{u}}{\partial s_{\mathrm{N}}^{u}}\mathcal{L}_{I_{\mathrm{N}}}\left(\frac{s_{\mathrm{N}}(x_{\mathrm{L}})t}{G_{k}\theta}\right)\bigg]\mathrm{d}t,
\end{aligned}
\label{eq:tauN}
\vspace{-0.5cm}
\end{equation}
\begin{equation}\small
   \tau_{\mathrm{R}}=\frac{W_{\mathrm{R}}}{\ln{2}}\int_{0}^{\infty}\frac{1}{t+1}\mathbb{E}_{x_{\mathrm{R}}}\left[\mathcal{L}_{I_{\mathrm{R}}}\left(\frac{s_{\mathrm{R}}(x_{\mathrm{R}})t}{\theta}\right)\exp\left(-\frac{\sigma_{\mathrm{R}}^2 s_{\mathrm{R}}(x_{\mathrm{R}})t}{\theta}\right)\right]\mathrm{d}t,
   \label{eq:tauR}
\end{equation}
where $s_{\mathrm{L}}(x_{\mathrm{L}})$, $s_{\mathrm{N}}(x_{\mathrm{N}})$ and $s_{\mathrm{R}}(x_{\mathrm{R}})$ are given in Theorem~\ref{theorem:coverage}. $W_{\mathrm{T}}$ and $W_{\mathrm{R}}$ are the bandwidth used in the THz and the RF communications, respectively. $\mathcal{L}_{I_{\mathrm{L}}}(\cdot)$, $\mathcal{L}_{I_{\mathrm{N}}}(\cdot)$ and $\mathcal{L}_{I_{\mathrm{R}}}(\cdot)$ are the Laplace transforms of the interference for the three association scenarios and are provided in (\ref{eq:LIL}), (\ref{eq:LIN}) and (\ref{eq:LIR}), respectively.
\begin{IEEEproof}
\emph{See Appendix~\ref{app:rate}.}
\end{IEEEproof}
\label{theorem:rate}
\end{theorem}
\begin{table*}[t]
	\centering
	\caption{Simulation Parameters.}
	\vspace{-0.3cm}
	\begin{tabular}{c|c||c|c||c|c}
		\hline\hline
		\textbf{Parameter} & \textbf{Value} & \textbf{Parameter} & \textbf{Value}  & \textbf{Parameter} & \textbf{Value} \\\hline
		($P_{\mathrm{T}}$, $P_{\mathrm{R}}$) & $5$~dBm & ($f_{\mathrm{T}}$, $f_{\mathrm{R}}$) & $1.05$~THz, $2.1$~GHz & ($G_{\mathrm{T}}^{(\mathrm{max})}$, $G_{\mathrm{T}}^{(\mathrm{min})}$) & ($25$, $-10$)~dB \\\hline
		$r_{d}$ & $80$~m & ($W_{\mathrm{T}}$, $W_{\mathrm{R}}$) & $0.5$~GHz, $40$~MHz & ($\varphi_{\mathrm{T}}$, $\varphi_{\mathrm{U}}$) & ($10^{\circ}$, $33^{\circ}$) \\\hline
		$N_{\mathrm{A}}$ & $20$ & ($\alpha_{\mathrm{L}}$, $\alpha_{\mathrm{N}}$, $\alpha_{\mathrm{R}}$) & ($2$, $4$ , $2.7$) & ($\sigma_{\varepsilon_{\mathrm{T}}}$, $\sigma_{\varepsilon_{\mathrm{U}}}$) & $0^{\circ}$ \\\hline
		$\delta_{\mathrm{T}}$ & $0.8$ & ($m_{\mathrm{L}}$, $m_{\mathrm{N}}$) & ($3$ , $1$) & $\lambda_{\mathrm{B}}$ &  $0.3$~m$^{-1}$ \\\hline
		($v_{0}$, $h_{\mathrm{U}}$) & ($0$ , $1.4$)~m & ($\sigma_{\mathrm{T}}^{2}$, $\sigma_{\mathrm{R}}^{2}$) & $4\times 10^{-11}$ & ($h_{\mathrm{A}}$, $r_{\mathrm{B}}$, $h_{\mathrm{B}}$) & ($4.5$, $0.22$, $1.7$)~m \\\hline
		$k_{a}(f_{\mathrm{T}})$ & $0.07512$~m$^{-1}$ & ($G_{\mathrm{U}}^{(\mathrm{max})}$, $G_{\mathrm{U}}^{(\mathrm{min})}$) & ($15$, $-10$)~dB & $\theta$ &  $0$~dB\\\hline\hline
	\end{tabular}
	\label{table:simulation_parameters}
	\vspace{-1.4cm}
\end{table*}
\vspace{-0.7cm}
\section{Numerical Results and Discussions}\label{sec:results}
In this section, we study the performance of the proposed RF and THz coexisting network and validate the analytical derivations through Monte-Carlo simulations. Furthermore, we investigate the effects of different system parameters and devise useful insights. The analysis is focused on a reference UE located in the center ($\textbf{v}_{0}=(0,0,0)$) of a finite disk of radius $r_{d}=80$~m where a fixed number $N_{\mathrm{A}}=20$ of THz and RF APs are deployed. Unless stated otherwise, the used simulation parameters are provided in Table~\ref{table:simulation_parameters}.
\begin{figure}[t]
	\centering
	\begin{minipage}[t]{0.47\linewidth}
		\includegraphics[width=\linewidth]{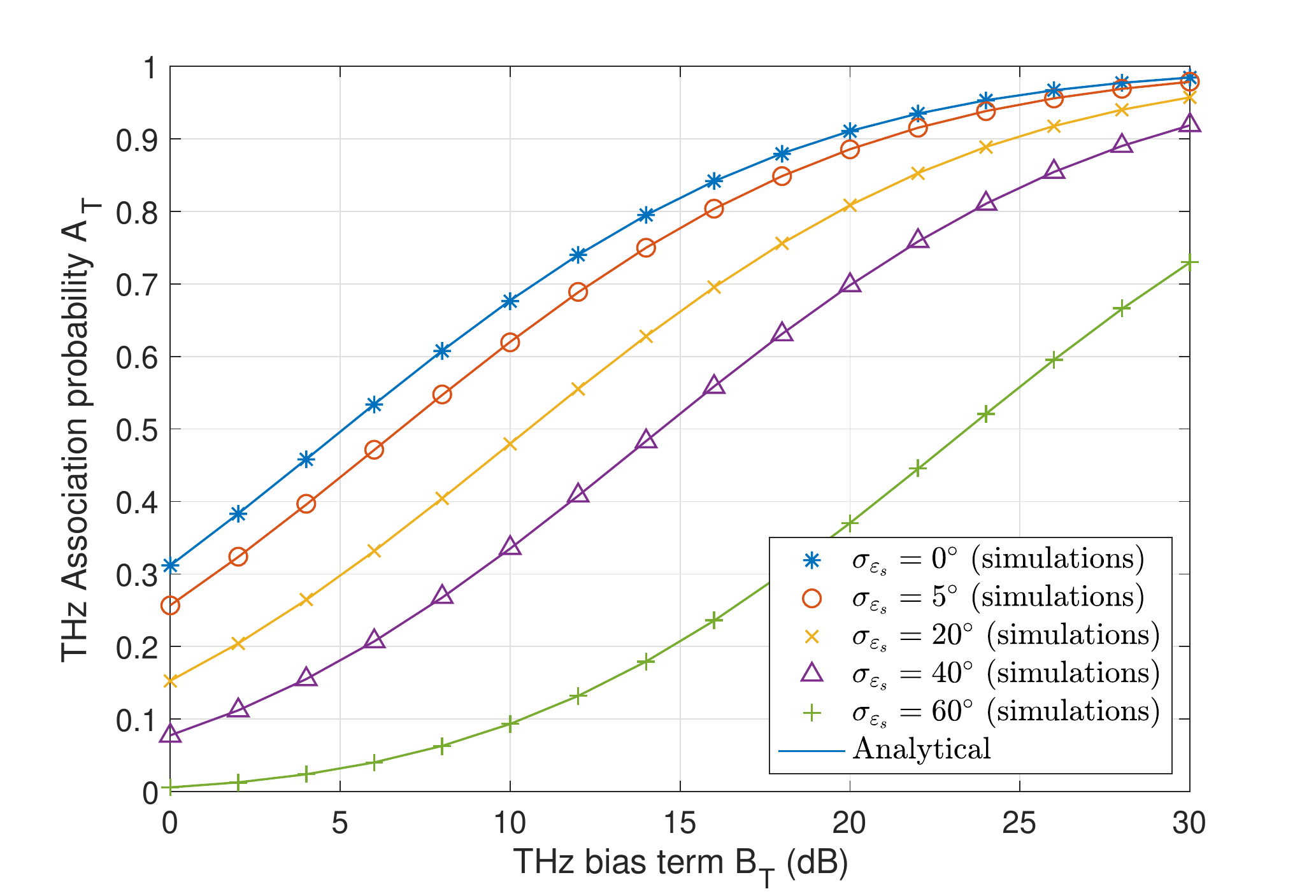}
		\vspace{-0.8cm}
		\caption{THz association probability as function of the bias factor $B_{\mathrm{T}}$ for different values of the misalignment error.}	
		\label{fig:AT_misalignment}
	\end{minipage}
     \begin{minipage}[t]{0.47\linewidth}
		\includegraphics[width=\linewidth]{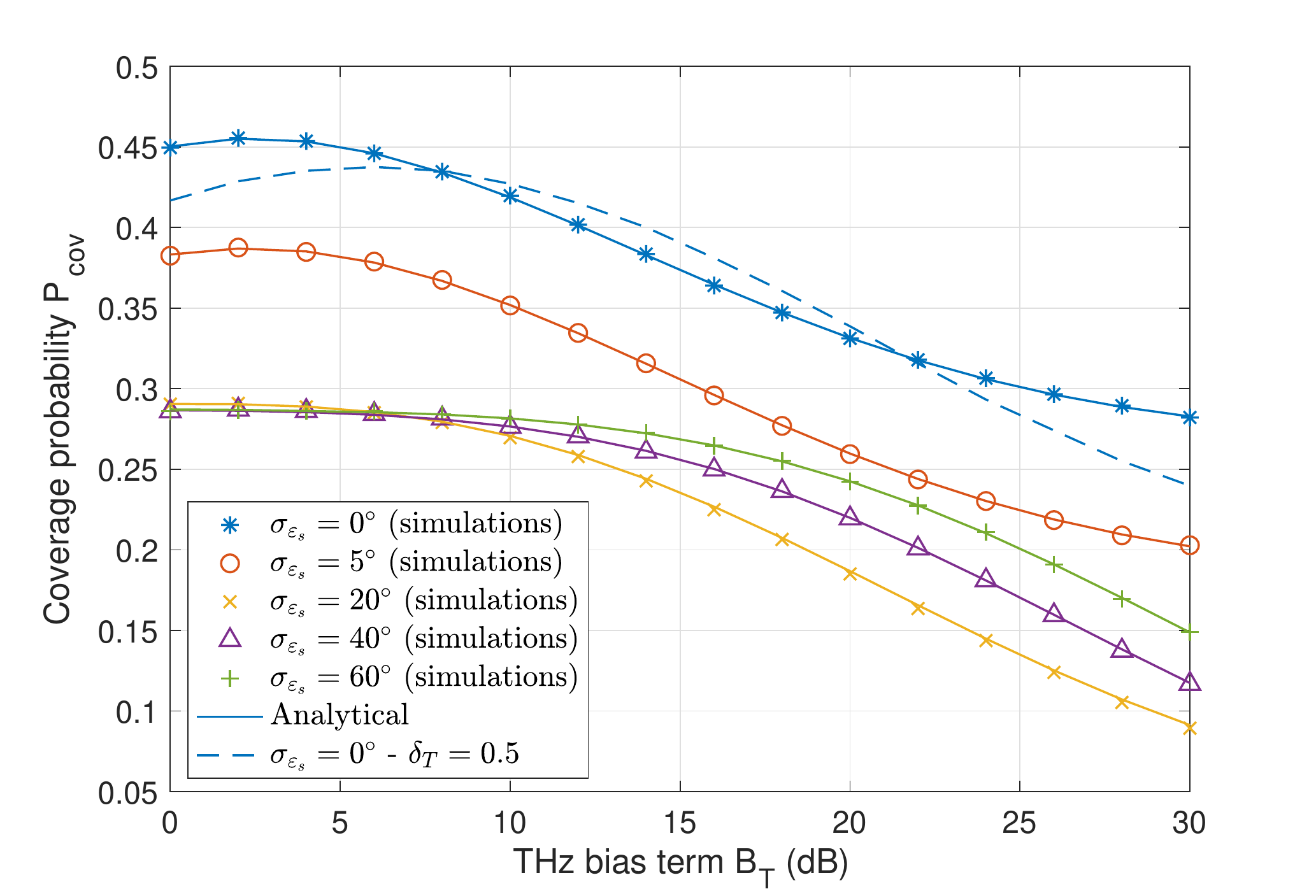}
		\vspace{-0.8cm}
		\caption{Coverage probability as function of the bias factor $B_{\mathrm{T}}$ for different values of the misalignment error and  $\delta_{\mathrm{T}}$.}	
		\label{fig:pcov_misalignment}
	\end{minipage}
	\vspace{-0.7cm}
\end{figure}

Fig.~\ref{fig:AT_misalignment} shows the simulation (markers) and analytical (solid lines) results of the THz association probability $A_{\mathrm{T}}=A_{\mathrm{L}}+A_{\mathrm{N}}$ as function of the bias term $B_{\mathrm{T}}$ for different values of the beam-steering error on the THz connection. We can see clearly that the analytical results match perfectly with the simulations, proving therefore the accuracy of the developed framework and the derived expressions in lemmas~\ref{lemma:AL} and~\ref{lemma:AN}. The same observation can be noted in Fig.~\ref{fig:pcov_misalignment} that shows the simulation and analytical results of the coverage probability as function of $B_{\mathrm{T}}$ for different values of $\sigma_{\varepsilon_s}$, $s\in \{\mathrm{T},\mathrm{U}\}$. Thus, the main analytical findings of this work, which are provided in Theorem~\ref{theorem:coverage}, are also validated. As expected, the THz association probability increases when the THz bias term increases, allowing as result to offload more UEs to THz APs. However, having a high misalignment error on the THz connection would limit such offloading as the UE will receive from the THz APs through side lobes only, thus with reduced power. 

An interesting trend can be noticed in~Fig.~\ref{fig:pcov_misalignment}; as the variance of the misalignment error increases, the coverage probability decreases. However, when the misalignment error reaches a certain level, the coverage probability starts to increase again. For high misalignment error levels, the THz-UE link quality deteriorates and the UE tends to associate more with the existing RF APs which are characterized with higher communication ranges. Thus, the overall coverage probability is improved. The low communication range of THz APs, which is mainly due to high absorption losses, limits the coverage probability that shows a slight improvement for low biasing values and faster degradation as $B_{\mathrm{T}}$ increases. Such behavior is clearly shown for $\delta_{\mathrm{T}}=0.5$ and the optimal bias to THz is larger with lower number of deployed THz APs. Thus, UEs are less encouraged to associate with the THz tier as it becomes more dense. Furthermore, the THz bias term should be optimized so as to maximize the coverage probability. 
\begin{figure}[t]
	\centering
	\begin{minipage}[t]{0.47\linewidth}
		\includegraphics[width=\linewidth,height=0.68\linewidth]{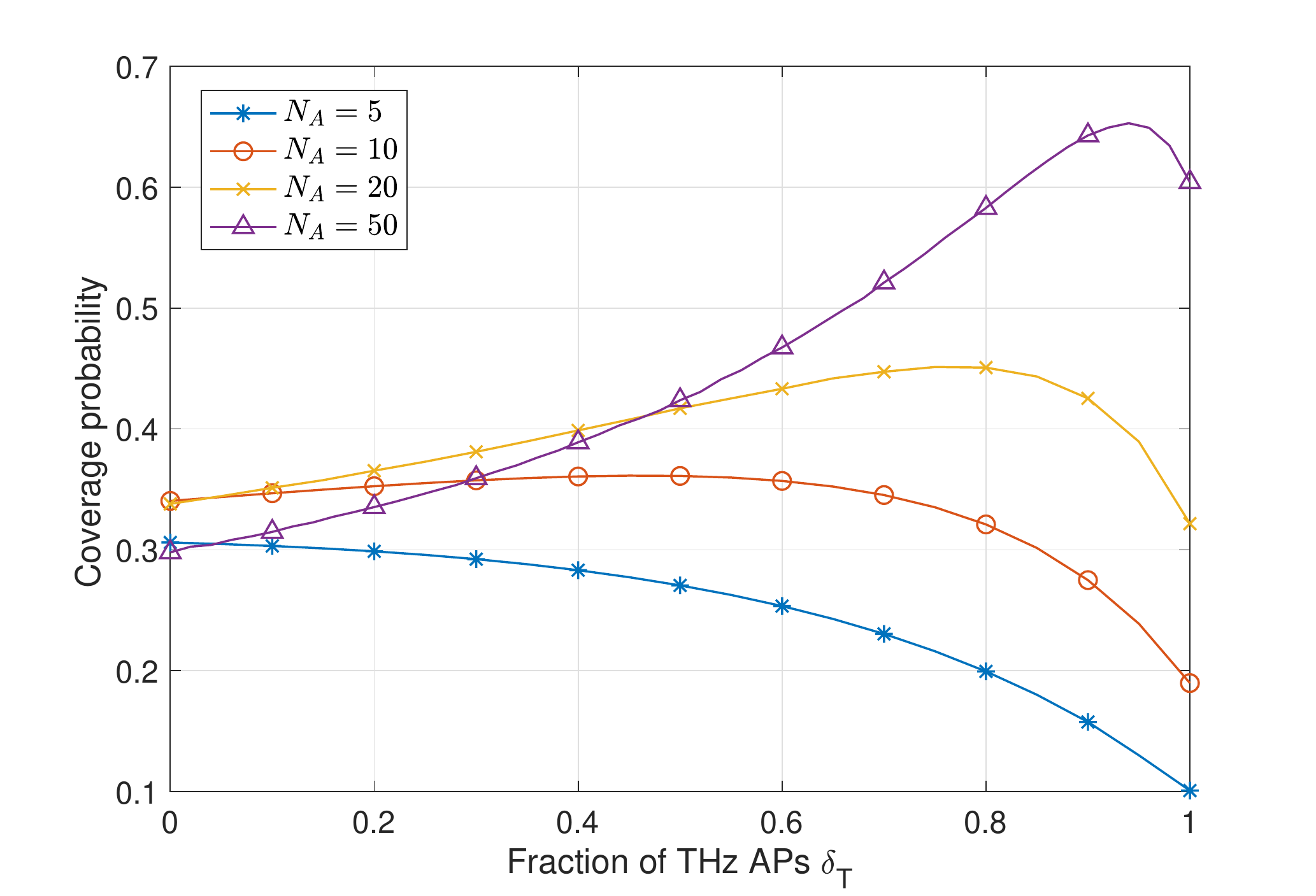}
		\vspace{-0.8cm}
		\caption{Coverage probability as function of $\delta_{\mathrm{T}}$ for different values of $N_{\mathrm{A}}$.}	
		\label{fig:pcov_deltaT_NA}
	\end{minipage}
	\begin{minipage}[t]{0.47\linewidth}
		\includegraphics[width=\linewidth,height=0.68\linewidth]{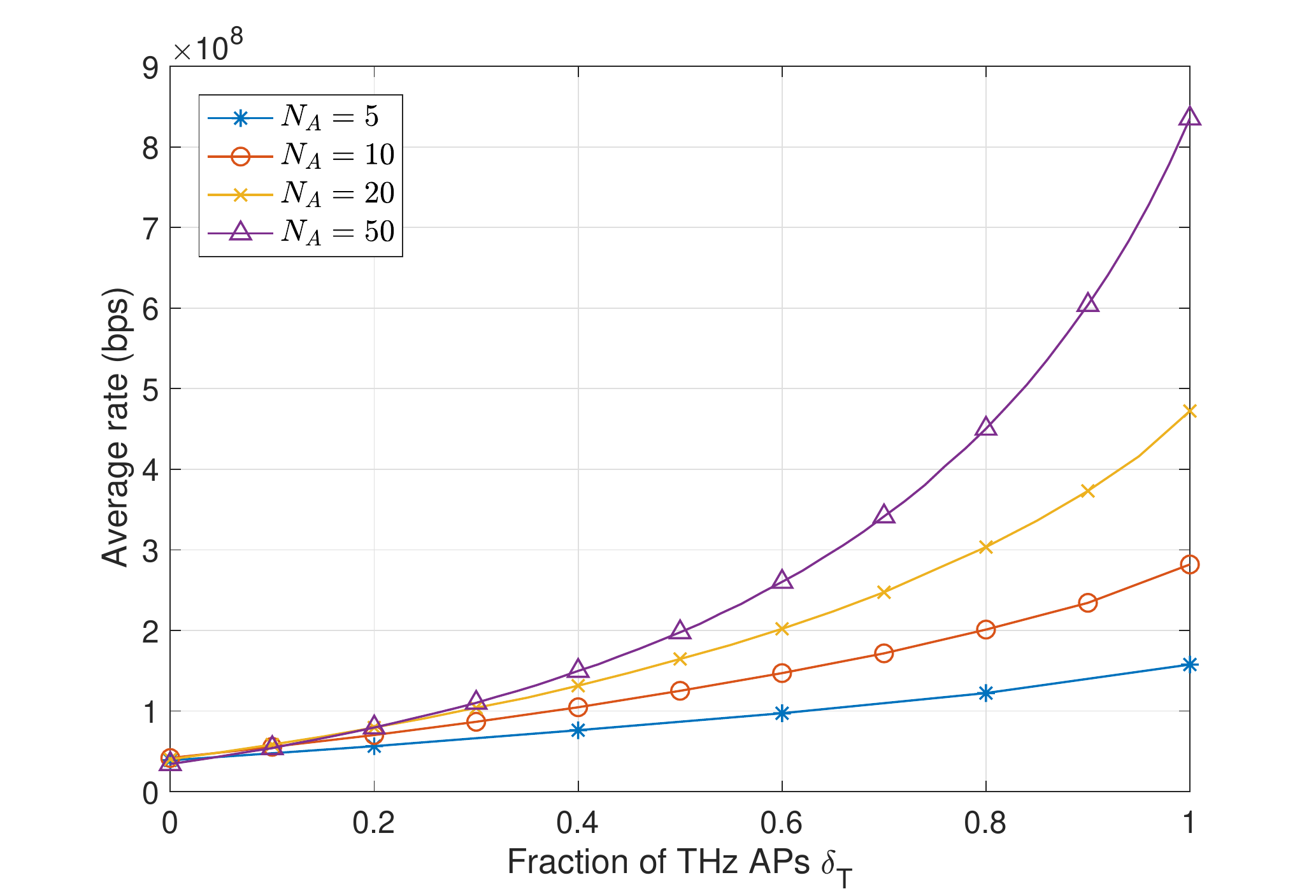}
		\vspace{-0.8cm}
		\caption{Average rate as function of $\delta_{\mathrm{T}}$ for different values of $N_{\mathrm{A}}$.}	
		\label{fig:rate_deltaT_NA}
	\end{minipage}
	\vspace{-0.7cm}
\end{figure}

In Fig.~\ref{fig:pcov_deltaT_NA}, we present the coverage probability versus the fraction of THz APs for different values of the total number of deployed THz and RF APs. For sparse deployments ($N_{\mathrm{A}}\leq 10$), increasing the fraction of THz APs results in the degradation of the coverage probability due to the limited communication range of THz APs and since the number of deployed APs is not sufficient to cover the considered finite area. For dense deployments, the coverage probability starts to increase as the fraction of THz APs increases. This happens up to a certain level after which adding more THz APs deteriorates the coverage probability. The initial improvement of the coverage probability is caused by the high gain directional antennas implemented at THz APs which will reduce the effective interference. However, if the majority of deployed APs are THz APs with limited communicated range, there is a higher chance that more UEs will fall into uncovered regions within the deployment area which will induce the degradation of the coverage probability for high $\delta_{\mathrm{T}}$ values. In addition, Fig.~\ref{fig:pcov_deltaT_NA} shows that, for low fractions of THz APs, densifying the network reduces the coverage probability which is dominated in this case by the behaviour of the RF tier and the high encountered interference levels from RF APs. On the other side, the impact of interference is mitigated as more THz APs with directional antennas and beam-steering capabilities are introduced.

Fig.~\ref{fig:rate_deltaT_NA} demonstrates the impact of the number of deployed APs and the fraction of THz APs on the average achievable rate in the RF and THz coexisting network. When compared to Fig.~\ref{fig:pcov_deltaT_NA}, a clear trade-off can be noticed; as the fraction of THz APs $\delta_{\mathrm{T}}$ increases, the average achievable rate significantly increases. Furthermore, densifying the network always improves the rate. This happens regardless of the coverage probability which decreases when THz APs dominate the network. Thus, the fraction of THz APs should be chosen carefully so as to optimize both coverage and rate performance. In fact, due to the low frequency band and limited bandwidth, sub-$6$~GHz technologies cannot satisfy very high demand for data rates. Such demand entails having small cells that operate at high frequencies such as THz. To this extend, the coexistence of THz and RF APs together can offer an attractive solution to meet the ever increasing need of ultra-high data rates while overcoming the limited coverage caused by the high absorption losses and high path loss in THz communications. 
\begin{figure}[t]
	\centering
	\begin{minipage}[t]{0.47\linewidth}
		\includegraphics[width=\linewidth,height=0.68\linewidth]{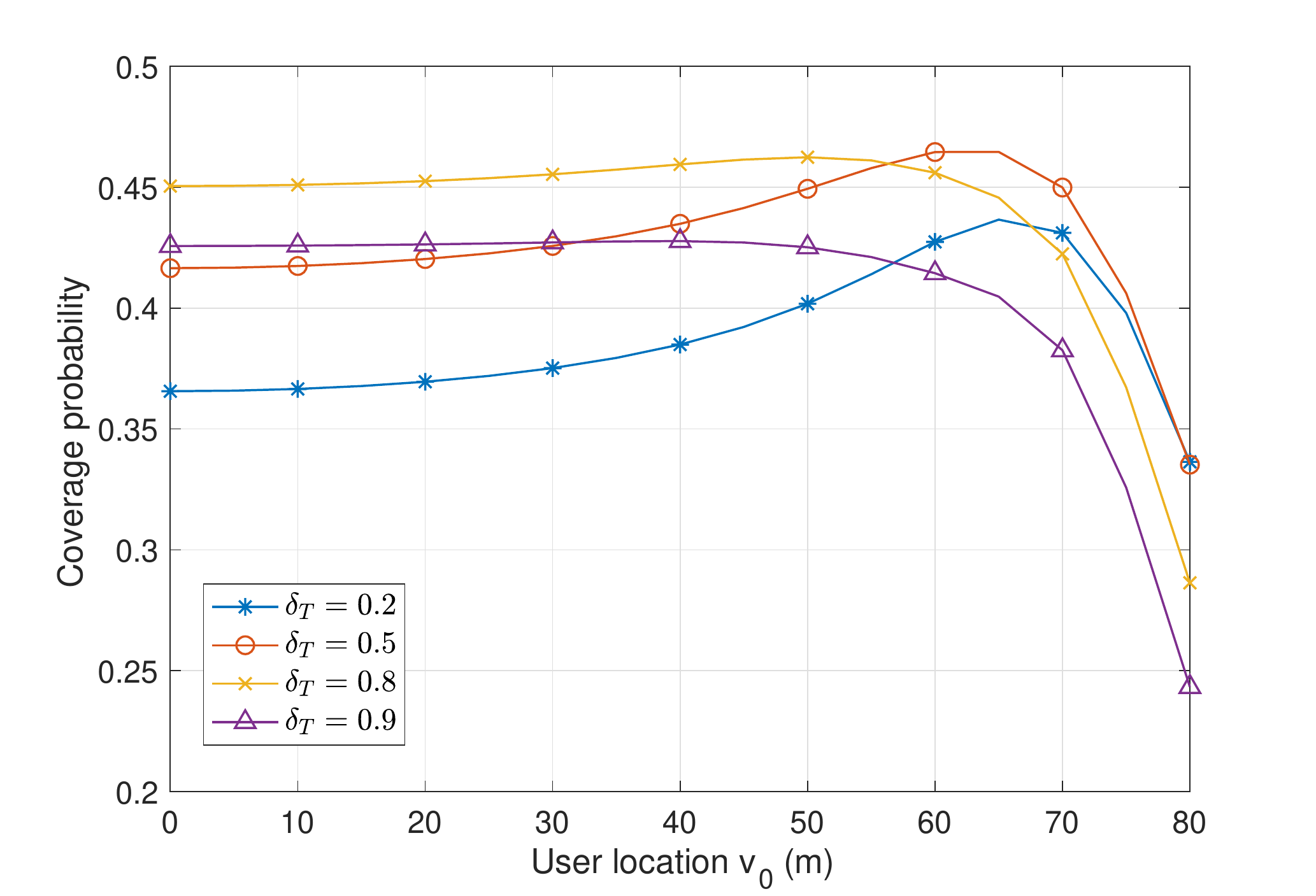}
		\vspace{-0.8cm}
		\caption{Coverage probability as function of the UE location for different values of $\delta_{\mathrm{T}}$.}	
		\label{fig:pcov_deltaT_x0}
	\end{minipage}
	\begin{minipage}[t]{0.47\linewidth}
		\includegraphics[width=\linewidth,height=0.68\linewidth]{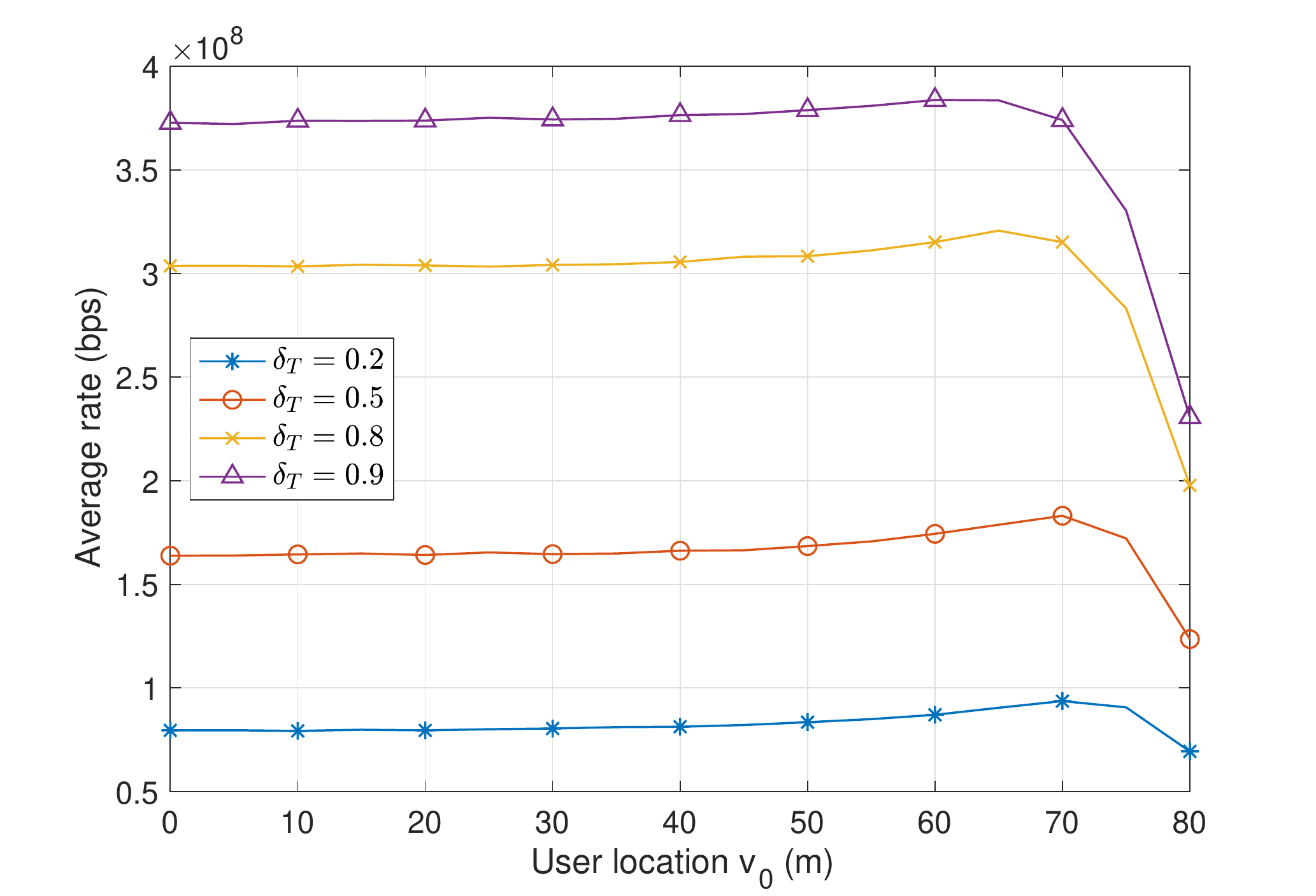}
		\vspace{-0.8cm}
		\caption{Average rate as function of the UE location for different values of $\delta_{\mathrm{T}}$.}	
		\label{fig:rate_deltaT_x0}
	\end{minipage}
	\vspace{-0.7cm}
\end{figure}

In Fig.~\ref{fig:pcov_deltaT_x0}, we evaluate the impact of the UE location in the finite area on the coverage probability. As the UE moves away from the network center, the coverage probability increases slowly. Such behavior is better noted for low number of deployed THz APs. As the UE gets closer to the network edge, the coverage probability drops significantly. The initial increase in the coverage probability, when moving away from the center of the network, is caused by the reduction of the interference received at the UE due to the border effect of the finite area. To this extend, the impact of interference is more severe in RF communications compared to THz communications because of the added THz-specific losses that limit the communication range, in addition to the highly directional antennas used for THz. Thus, the improvement of the coverage probability as the UE approaches the network boundary is achieved when RF APs are dominating the network. The degradation of the coverage probability at the edge of the network is caused by the small likelihood to find a close by AP to serve the UE. Furthermore, central UEs show better coverage performance for high fraction of THz APs $\delta_{\mathrm{T}}$, while edge UEs present higher coverage probability for low values of $\delta_{\mathrm{T}}$. The main reason for such behavior is the limited communication range of THz APs, which makes finding a good THz AP to serve the UE on the edge more challenging.  

The impact of the UE location on the average rate is captured in Fig.~\ref{fig:rate_deltaT_x0}. Similarly to the coverage probability, the maximum average rate value is achieved when the UE is close to the boundary of the finite area. However, increasing the fraction of THz APs always improve the average rate. The results in Fig.~\ref{fig:pcov_deltaT_x0} and Fig.~\ref{fig:rate_deltaT_x0} highlight clearly the importance of taking into consideration the UE location when choosing the best fraction of THz APs to be deployed in a finite area so as to optimize both the coverage and the achievable rate.   
\vspace{-0.5cm}

\section{Conclusion}\label{sec:conclusion}
This paper investigates the coverage and rate performance of a coexisting RF and THz finite network using tools from stochastic geometry. Furthermore, we highlight the impact of different system parameters such as the fraction of THz APs, the THz beam-steering error, and the UE location. Based on the developed framework, we derive tractable expressions for the association probabilities with the THz and the RF tiers, the serving distance distributions, the conditional coverage probabilities, and the average achievable rate. The obtained results reveal that densifying the network with THz APs can improve the rate but negatively affect the coverage probability. Furthermore, a clear trade-off exists between the fraction of THz APs in the network and the bias to the THz tier. Thus, deploying RF and THz APs in a finite area should be carefully planned to achieve ultra-high rates while maintaining sufficient coverage. This work can be extended by using a more sophisticated and realistic blockage model that accounts for the impact of walls and furniture in indoor environments. Furthermore, the impact of intelligent reflective surfaces in hybrid RF/THz networks can be investigated.
\vspace{-0.5cm}
\appendix
\subsection{Proof of Lemma~\ref{lemma:AL}}\label{app:AL}
By referring to the association rule, the UE can associate with the AP that offers the strongest average BRSP. The corresponding association probabilities are derived by adopting a similar approach to~\cite{Wang}. Consider an arbitrary AP placed at a distance $r$ from the reference UE, this AP is the serving LOS THz AP when three events are simultaneously fulfilled:
\begin{itemize}
	\item The AP located at a distance $r$ from the UE is a THz AP with LOS connection. Given that $\delta_{\mathrm{T}}$ is the fraction of THz APs and $\kappa_{\mathrm{L}}(\cdot)$ is the LOS probability, the probability that this event occurs is $\delta_{\mathrm{T}}\kappa_{\mathrm{L}}(r)$.
	\item For the $(\delta_{\mathrm{T}}N_{\mathrm{A}}-1)$ remaining THz APs, each AP is either a LOS AP located at a greater distance than $r$ or a NLOS AP located at a greater distance than the exclusion region $E_{\mathrm{LN}}(r)$ given in (\ref{eq:ELN}). The probabilities of occurrence of these two cases can be obtained as $\int_{r}^{z_{p}}f_{\mathrm{Z}}(z)\kappa_{\mathrm{L}}(z)\mathrm{d}z$ and $\int_{E_{\mathrm{LN}}(r)}^{z_{p}}f_{\mathrm{Z}}(z)\kappa_{\mathrm{N}}(z)\mathrm{d}z$, respectively, where $f_{\mathrm{Z}}(\cdot)$ is given in (\ref{eq:fw}) and denotes the probability density function (PDF) of the distance from a random AP to the UE and $\kappa_{\mathrm{N}}(\cdot)$ is the NLOS probability. As the two events are mutually exclusive and the remaining $(\delta_{\mathrm{T}}N_{\mathrm{A}}-1)$ THz APs are independent and identically distributed (i.i.d) after conditioning on the location of the reference UE, the probability of achieving this condition is given as $\left(\int_{r}^{z_{p}}f_{\mathrm{Z}}(z)\kappa_{\mathrm{L}}(z)\mathrm{d}z+\int_{E_{\mathrm{LN}}(r)}^{z_{p}}f_{\mathrm{Z}}(z)\kappa_{\mathrm{N}}(z)\mathrm{d}z\right)^{\delta_{\mathrm{T}}N_{\mathrm{A}}-1}$
	\item For any of the remaining $(1-\delta_{\mathrm{T}})N_{\mathrm{A}}$ RF APs, it should be located further than the exclusion region $E_{\mathrm{LR}}(r)$ given in (\ref{eq:ELR}). Such event can occur with probability $\int_{E_{\mathrm{LR}}(r)}^{z_{p}}f_{\mathrm{Z}}(z)\mathrm{d}z$. As the $(1-\delta_{\mathrm{T}})N_{\mathrm{A}}$ RF APs are i.i.d after conditioning of the location of the UE, the probability of this condition is given as $\left(\int_{E_{\mathrm{LR}}(r)}^{z_{p}}f_{\mathrm{Z}}(z)\mathrm{d}z\right)^{(1-\delta_{\mathrm{T}})N_{\mathrm{A}}}$.
\end{itemize}
 As the three conditions above are independent, we can derive the probability that the LOS THz AP at distance $r$ is the serving AP as the multiplication of the probabilities of the three events. Finally, there are $N_{\mathrm{A}}$ ways of choosing an AP from the BPP set of APs $\Phi_{\mathrm{A}}$. As result, the probability that the UE is associated with a LOS THz AP located at distance $r$ is given by:
 \begin{equation}\small
 	N_{\mathrm{A}}\delta_{\mathrm{T}}\kappa_{\mathrm{L}}(r)\left(\int_{r}^{z_{p}}f_{\mathrm{Z}}(z)\kappa_{\mathrm{L}}(z)\mathrm{d}z+\int_{E_{\mathrm{LN}}(r)}^{z_{p}}f_{\mathrm{Z}}(z)\kappa_{\mathrm{N}}(z)\mathrm{d}z\right)^{\delta_{\mathrm{T}}N_{\mathrm{A}}-1}\left(\int_{E_{\mathrm{LR}}(r)}^{z_{p}}f_{\mathrm{Z}}(z)\mathrm{d}z\right)^{(1-\delta_{\mathrm{T}})N_{\mathrm{A}}}.
 \end{equation} 
The final expression of the association probability with a LOS THz AP $A_{\mathrm{L}}$ given in (\ref{eq:AL}) can be derived by integrating over $z_{l}\leq r \leq z_{p}$.
\vspace{-0.4cm}
\subsection{Proof of Lemma~\ref{lemma:xL}}\label{app:servingdistance}
\vspace{-0.1cm}
The distribution of the distance separating the UE from the serving LOS THz AP $x_{\mathrm{L}}$ is equivalent to the distribution of $d_{\mathrm{L}}$, where $d_{\mathrm{L}}$ is the distance to the closest LOS THz AP, given that the UE associates with a LOS THz AP (i.e. given that the event $C_{\mathrm{L}}$ occurs). Thus, the complementary cumulative distribution function (CCDF) of $X_{\mathrm{L}}$ can be obtained as
\begin{equation}\small
	\bar{F}_{X_{\mathrm{L}}}(x_{\mathrm{L}})=\mathbb{P}\left[d_{\mathrm{L}}>x_{\mathrm{L}}|C_{\mathrm{L}}\right]=\frac{\mathbb{P}\left[d_{\mathrm{L}}>x_{\mathrm{L}}, C_{\mathrm{L}}\right]}{\mathbb{P}\left[C_{\mathrm{L}}\right]},
	\label{eq:barFxL}
\end{equation}
where $\mathbb{P}\left[C_{\mathrm{L}}\right]=A_{\mathrm{L}}$ is given in (\ref{eq:AL}). For the case when the UE chooses to associate with a LOS THz APs at distance $r$, any of the remaining APs is either a LOS THz AP located at a greater distance than $r$, a NLOS THz AP located at a greater distance than $E_{\mathrm{LN}}(r)$ or an RF AP located further than $E_{\mathrm{LR}}(r)$. As the $N_{\mathrm{A}}$ APs are i.i.d after conditioning on the location of the UE, and given that a fraction $\delta_{\mathrm{T}}$ of these APs are THz APs, the numerator of (\ref{eq:barFxL}) is given as
\begin{equation}\small
\begin{aligned}
	&\mathbb{P}\left[d_{\mathrm{L}}>x_{\mathrm{L}}, C_{\mathrm{L}}\right]=\delta_{\mathrm{T}} N_{\mathrm{A}}\\
	&\times\int_{x_{\mathrm{L}}}^{z_{p}}f_{\mathrm{Z}}(r)\kappa_{\mathrm{L}}(r)\left(\int_{E_{\mathrm{LR}(r)}}^{z_{p}}f_{\mathrm{Z}}(z)\mathrm{d}z\right)^{(1-\delta_{\mathrm{T}})N_{\mathrm{A}}}\left(\int_{r}^{z_{p}}f_{\mathrm{Z}}(z)\kappa_{\mathrm{L}}(z)\mathrm{d}z+\int_{E_{\mathrm{LN}}(r)}^{z_{p}}f_{\mathrm{Z}}(z)\kappa_{\mathrm{N}}(z)\mathrm{d}z\right)^{\delta_{\mathrm{T}} N_{\mathrm{A}}-1}\mathrm{d}r,
\end{aligned}
\end{equation}
where $f_{\mathrm{Z}}(\cdot)$, $\kappa_{\mathrm{L}}(\cdot)$ and $\kappa_{\mathrm{N}}(\cdot)$ are given in (\ref{eq:fw}), (\ref{eq:PL}) and (\ref{eq:PN}), respectively. $E_{\mathrm{LR}}(\cdot)$ and $E_{\mathrm{LN}}(\cdot)$ are the exclusion regions on the locations of the remaining APs and are given in (\ref{eq:ELR}) and (\ref{eq:ELN}), respectively. The cumulative distribution function (CDF) of $X_{\mathrm{L}}$ is $F_{X_{\mathrm{L}}}(x_{\mathrm{L}})=1-\bar{F}_{X_{\mathrm{L}}}(x_{\mathrm{L}})$ and the PDF $f_{X_\mathrm{L}}(x_{\mathrm{L}})=\frac{\mathrm{d}F_{X_{\mathrm{L}}}(x_{\mathrm{L}})}{\mathrm{d}x_{\mathrm{L}}}$ is given as in (\ref{eq:fxL}).
\vspace{-0.4cm}
\subsection{Proof of Theorem~\ref{theorem:coverage}}\label{app:coverage}
The conditional coverage probability $P_{cov,\mathrm{L}}$ given that the UE associates with a LOS THz AP located at $x_{\mathrm{L}}$ (i.e. the event $C_{\mathrm{L}}$ occurs) is derived as
\begin{equation}\small
	\begin{aligned}
	P_{cov,\mathrm{L}}&=\mathbb{P}\left[\mathrm{SINR}\geq\theta|C_{\mathrm{L}}\right]=\mathbb{P}\left[\mathrm{SINR}_{\mathrm{L}}\geq \theta\right]=\mathrm{P}\left[\frac{P_{\mathrm{T}} \gamma_{\mathrm{T}} G_{\mathrm{T},0} e^{-k_a(f_{\mathrm{T}})x_{\mathrm{L}}}x_{\mathrm{L}}^{-\alpha_{\mathrm{L}}}\chi_{\mathrm{L},0}}{I_{\mathrm{L}}+\sigma_{\mathrm{T}}^2}\geq\theta\right]\\
	&\stackrel{(a)}{=}\mathbb{E}_{G_{\mathrm{T},0},x_{\mathrm{L}},I_{\mathrm{L}}}\left[\mathbb{P}\left[\chi_{\mathrm{L},0}\geq \frac{\theta(I_{\mathrm{L}}+\sigma_{\mathrm{T}}^2)}{P_{\mathrm{T}} \gamma_{\mathrm{T}} G_{\mathrm{T},0} e^{-k_a(f_{\mathrm{T}})x_{\mathrm{L}}}x_{\mathrm{L}}^{-\alpha_{\mathrm{L}}}}\right]\right]\\
	&\stackrel{(b)}{=}\sum_{k=1}^{4}p_{k,0}\mathbb{E}_{x_{\mathrm{L}},I_{\mathrm{L}}}\left[\sum_{q=0}^{m_{\mathrm{L}}-1}\frac{1}{q!}\left(\frac{m_{\mathrm{L}}\theta(I_{\mathrm{L}}+\sigma_{\mathrm{T}}^2)}{P_{\mathrm{T}} \gamma_{\mathrm{T}} G_{k} e^{-k_a(f_{\mathrm{T}})x_{\mathrm{L}}}x_{\mathrm{L}}^{-\alpha_{\mathrm{L}}}}\right)^{q}\exp\left(-\frac{m_{\mathrm{L}}\theta(I_{\mathrm{L}}+\sigma_{\mathrm{T}}^2)}{P_{\mathrm{T}} \gamma_{\mathrm{T}} G_{k} e^{-k_a(f_{\mathrm{T}})x_{\mathrm{L}}}x_{\mathrm{L}}^{-\alpha_{\mathrm{L}}}}\right)\right]\\
	&\stackrel{(c)}{=}\sum_{k=1}^{4}p_{k,0}\mathbb{E}_{x_{\mathrm{L}},I_{\mathrm{L}}}\left[\sum_{q=0}^{m_{\mathrm{L}}-1}\frac{(s_{\mathrm{L}}(x_{\mathrm{L}}))^q}{q!}\left(\frac{I_{\mathrm{L}}+\sigma_{\mathrm{T}}^{2}}{G_{k}}\right)^{q}\exp\left(-\frac{I_{\mathrm{L}}+\sigma_{\mathrm{T}}^{2}}{G_{k}}s_{\mathrm{L}}(x_{\mathrm{L}})\right)\right]\\
	&\stackrel{(d)}{=}\sum_{k=1}^{4}p_{k,0}\mathbb{E}_{x_{\mathrm{L}}}\left[\sum_{q=0}^{m_{\mathrm{L}}-1}\frac{(s_{\mathrm{L}}(x_{\mathrm{L}}))^q}{q!}\left[\frac{\partial^{q}}{\partial s_{\mathrm{L}}^{q}}\exp\left(-\frac{\sigma_{\mathrm{T}}^{2}s_{\mathrm{L}}(x_{\mathrm{L}})}{G_{\mathrm{k}}}\right)\mathcal{L}_{I_{\mathrm{L}}}\left(\frac{s_{\mathrm{L}}(x_{\mathrm{L}})}{G_{k}}\right)\right]\right],
	\end{aligned}
\end{equation}
where (a) is obtained by averaging the conditional coverage probability over $\{G_{\mathrm{T},0},x_{\mathrm{L}},I_{\mathrm{L}}\}$ and from exploiting the independence between them. (b) follows from the CCDF of the Nakagami small scale fading gain $\chi_{\mathrm{L},0}$ given in (\ref{eq:barFxL}) and from averaging over the discrete random variable $G_{\mathrm{T},0}$ corresponding to the directionality gain of the desired link whose PMF is given in Table~\ref{table:antenna}. (c) is obtained from denoting $s_{\mathrm{L}}(x_{\mathrm{L}})=\frac{m_{\mathrm{L}}\theta e^{-k_{a}(f_{\mathrm{T}})x_{\mathrm{L}}}x_{\mathrm{L}}^{-\alpha_{\mathrm{L}}}}{P_{\mathrm{T}}\gamma_{\mathrm{T}}}$ and (d) is obtained from the partial derivative expression of the exponential term and from the Laplace transform definition $\mathcal{L}_{I_{\mathrm{L}}}(s)=\mathbb{E}_{I_{\mathrm{L}}}\left[e^{-s I_{\mathrm{L}}}\right]$. The final expression of $P_{cov,\mathrm{L}}$ given in (\ref{eq:PcovL}) is obtained from applying the expression:
\vspace{-0.3cm}
\begin{equation}\small
	\frac{\partial^{q}}{\partial x^{q}}f(x)g(x)=\sum_{u=0}^{q}{q\choose u} \frac{\partial^{u}}{\partial x^{u}}f(x)\frac{\partial^{q-u}}{\partial x^{q-u}}g(x).
\end{equation}
The conditional coverage probability $P_{cov,\mathrm{N}}$ when the UE is associated with a NLOS THz AP located at $x_{\mathrm{N}}$ (i.e. the event $C_{\mathrm{N}}$ occurs) is obtained following the same procedure as $P_{cov,\mathrm{L}}$ and is given in (\ref{eq:PcovN}). Finally, the conditional coverage probability $P_{cov,\mathrm{R}}$ when the UE associates with an RF AP located at $x_{\mathrm{R}}$ (i.e. the event $C_{\mathrm{R}}$ occurs) is
\begin{equation}\small
\begin{aligned}
	&P_{cov,\mathrm{R}}=\mathbb{P}\left[\mathrm{SINR}\geq\theta | C_{\mathrm{R}}\right]=\mathbb{P}\left[\mathrm{SINR}_{\mathrm{R}}\geq\theta\right]=\mathbb{P}\left[\frac{P_{\mathrm{R}} \gamma_{\mathrm{R}} x_{\mathrm{R}}^{-\alpha_{\mathrm{R}}}\chi_{\mathrm{R},0}}{I_{\mathrm{R}}+\sigma_{\mathrm{R}}^2}\geq\theta\right]\\
	&\stackrel{(a)}{=}\mathbb{E}_{x_{\mathrm{R}},I_{\mathrm{R}}}\left[\mathbb{P}\left[\chi_{\mathrm{R},0}\geq\frac{\theta(I_{\mathrm{R}}+\sigma_{\mathrm{R}}^2)}{P_{\mathrm{R}} \gamma_{\mathrm{R}} x_{\mathrm{R}}^{-\alpha_{\mathrm{R}}}}\right]\right]\stackrel{(b)}{=}\mathbb{E}_{x_{\mathrm{R}}}\left[\exp\left(-\frac{\theta\sigma_{\mathrm{R}}^2}{P_{\mathrm{R}} \gamma_{\mathrm{R}} x_{\mathrm{R}}^{-\alpha_{\mathrm{R}}}}\right)\mathbb{E}_{I_{\mathrm{R}}}\left[\exp\left(-\frac{\theta I_{\mathrm{R}}}{P_{\mathrm{R}} \gamma_{\mathrm{R}} x_{\mathrm{R}}^{-\alpha_{\mathrm{R}}}}\right)\right]\right],
\end{aligned}
\end{equation}
where (a) follows from averaging over the independent random variables $\{x_{\mathrm{R}},I_{\mathrm{R}}\}$ and (b) from the CCDF of the exponential small scale fading $\chi_{\mathrm{R},0}$. The expression given in (\ref{eq:PcovR}) is obtained by denoting $s_{\mathrm{R}}(x_{\mathrm{R}})=\frac{\theta}{P_{\mathrm{R}}\gamma_{\mathrm{R}}x_{\mathrm{R}}^{-\alpha_{\mathrm{R}}}}$ and from the definition of the Laplace transform of $I_{\mathrm{R}}$.
\vspace{-0.5cm}
\subsection{Proof of Lemma~\ref{lemma:LIL}}\label{app:LIL}
To derive the Laplace transform of the interference $I_{\mathrm{L}}$ from the THz APs in the case when the UE is associated with a LOS THz AP placed at a distance $x_{\mathrm{L}}$ from the UE, we refer to a similar procedure to~\cite{Wang}. Note here that $I_{\mathrm{L}}$ includes the interference from both the LOS and NLOS THz APs except the serving AP and is given in (\ref{eq:IL}). $I_{\mathrm{L}}$ can also be expressed as $I_{\mathrm{L}}=\sum_{i=1}^{\delta_{\mathrm{T}}N_{\mathrm{A}}-1}I_{\mathrm{L},x_{i}}$, where $I_{\mathrm{L},x_{i}}$ is the interference from the THz AP located at $\textbf{x}_{i}$. For any of the $(\delta_{\mathrm{T}} N_{\mathrm{A}}-1)$ interfering THz APs, it can be a LOS THz AP located at a greater than $x_{\mathrm{L}}$ or a NLOS THz AP located at a greater distance than $E_{\mathrm{LN}}(x_{\mathrm{L}})$. The probabilities of occurrence of these two events are $\frac{\int_{x_{\mathrm{L}}}^{z_{p}}f_{\mathrm{Z}}(z)\kappa_{\mathrm{L}}(z)\mathrm{d}z}{\int_{x_{\mathrm{L}}}^{z_{p}}f_{\mathrm{Z}}(z)\kappa_{\mathrm{L}}(z)\mathrm{d}z+\int_{E_{\mathrm{LN}}(x_{\mathrm{L}})}^{z_{p}}f_{\mathrm{Z}}(z)\kappa_{\mathrm{N}}(z)\mathrm{d}z}$ and  $\frac{\int_{E_{\mathrm{LN}}(x_{\mathrm{L}})}^{z_{p}}f_{\mathrm{Z}}(z)\kappa_{\mathrm{N}}(z)\mathrm{d}z}{\int_{x_{\mathrm{L}}}^{z_{p}}f_{\mathrm{Z}}(z)\kappa_{\mathrm{L}}(z)\mathrm{d}z+\int_{E_{\mathrm{LN}}(x_{\mathrm{L}})}^{z_{p}}f_{\mathrm{Z}}(z)\kappa_{\mathrm{N}}(z)\mathrm{d}z}$, respectively. The Laplace transform $\mathcal{L}_{I_{\mathrm{L}}}(s)$ is given as
\begin{equation}\small
\begin{aligned}
	\mathcal{L}_{I_{\mathrm{L}}}(s)&=\mathbb{E}_{I_{\mathrm{L}}}\left[e^{-s I_{\mathrm{L}}}\right]\\&=\mathbb{E}_{I_{\mathrm{L}}}\left[\exp\left(-s\sum_{i=1}^{\delta_{\mathrm{T}}N_{\mathrm{A}}-1}I_{\mathrm{L},x_{i}}\right)\right]
	\stackrel{(a)}{=}\prod_{i=1}^{\delta_{\mathrm{T}}N_{\mathrm{A}}-1}\mathbb{E}_{I_{\mathrm{L},x_{i}}}\left[\exp\left(-s I_{\mathrm{L},x_{i}}\right)\right]=\left(\mathbb{E}_{I_{\mathrm{L},x_{i}}}\left[\exp\left(-s I_{\mathrm{L},x_{i}}\right)\right]\right)^{\delta_{\mathrm{T}}N_{\mathrm{A}}-1},
\end{aligned}
\label{eq:LIL_proof}
\end{equation} 
where (a) is induced from the i.i.d distribution of the small scale fading gains and from their independence of the interferers distances and the directionality gains in the interference expression. The expectation term $\mathbb{E}_{I_{\mathrm{L},x_{i}}}\left[\exp\left(-s I_{\mathrm{L},x_{i}}\right)\right]$ can be calculated as
\begin{equation}\small
\begin{aligned}
	&\mathbb{E}_{I_{\mathrm{L},x_{i}}}\left[\exp\left(-s I_{\mathrm{L},x_{i}}\right)\right]=\frac{\int_{x_{\mathrm{L}}}^{z_{p}}f_{\mathrm{Z}}(z)\kappa_{\mathrm{L}}(z)\mathrm{d}z}{\int_{x_{\mathrm{L}}}^{z_{p}}f_{\mathrm{Z}}(z)\kappa_{\mathrm{L}}(z)\mathrm{d}z+\int_{E_{\mathrm{LN}}(x_{\mathrm{L}})}^{z_{p}}f_{\mathrm{Z}}(z)\kappa_{\mathrm{N}}(z)\mathrm{d}z}\mathbb{E}_{P_{\mathrm{L},x_{i}}^{r}}\left[\exp\left(-s P_{\mathrm{L},x_{i}}^{r}\right)\right]\\
	&+\frac{\int_{E_{\mathrm{LN}}(x_{\mathrm{L}})}^{z_{p}}f_{\mathrm{Z}}(z)\kappa_{\mathrm{N}}(z)\mathrm{d}z}{\int_{x_{\mathrm{L}}}^{z_{p}}f_{\mathrm{Z}}(z)\kappa_{\mathrm{L}}(z)\mathrm{d}z+\int_{E_{\mathrm{LN}}(x_{\mathrm{L}})}^{z_{p}}f_{\mathrm{Z}}(z)\kappa_{\mathrm{N}}(z)\mathrm{d}z}\mathbb{E}_{P_{\mathrm{N},x_{i}}^{r}}\left[\exp\left(-s P_{\mathrm{N},x_{i}}^{r}\right)\right],
\end{aligned}
\label{eq:EIL_proof}
\end{equation}
where $P_{\mathrm{L},x_{i}}^{r}$ and $P_{\mathrm{N},x_{i}}^{r}$ are the received powers from the interfering LOS THz AP at $\textbf{x}_{i}$ and the NLOS THz AP at $\textbf{x}_{i}$ given in Section~\ref{sec:THz_channel_model}. $\mathbb{E}_{P_{\mathrm{L},x_{i}}^{r}}\left[\exp\left(-s P_{\mathrm{L},x_{i}}^{r}\right)\right]$ can be obtained as 
\begin{equation}\small
\begin{aligned}
\mathbb{E}_{P_{\mathrm{L},x_{i}}^{r}}\left[\exp\left(-s P_{\mathrm{L},x_{i}}^{r}\right)\right]&\stackrel{(a)}{=}\mathbb{E}_{G_{\mathrm{T}},\chi_{\mathrm{L}},d_{\mathrm{L}}}\left[\exp\left(-s P_{\mathrm{T}}\gamma_{\mathrm{T}}G_{\mathrm{T}}e^{-k_a(f_{\mathrm{T}})d_{\mathrm{L}}}d_{\mathrm{L}}^{-\alpha_{\mathrm{L}}}\chi_{\mathrm{L}}\right)\right]\\
&\stackrel{(b)}{=}\sum_{k=1}^{4}p_{k}\mathbb{E}_{d_{\mathrm{L}}}\left[\left(1+\frac{s P_{\mathrm{T}}\gamma_{\mathrm{T}}G_{k}e^{k_{a}(f_{\mathrm{T}})d_{\mathrm{L}}}d_{\mathrm{L}}^{-\alpha_{\mathrm{L}}}}{m_{\mathrm{L}}}\right)^{-m_{\mathrm{L}}}\right]\\
&\stackrel{(c)}{=}\sum_{k=1}^{4}p_{k}\int_{x_{\mathrm{L}}}^{z_{p}}\left(1+\frac{s P_{\mathrm{T}}\gamma_{\mathrm{T}}G_{k}e^{k_{a}(f_{\mathrm{T}})y}y^{-\alpha_{\mathrm{L}}}}{m_{\mathrm{L}}}\right)^{-m_{\mathrm{L}}}f_{Y_{\mathrm{L}}}(y,x_{\mathrm{L}})\mathrm{d}y,
\end{aligned}
\label{eq:EPLr_proof}
\end{equation}
where (a) is obtained from replacing $P_{\mathrm{L},x_{i}}$ with its expression and omitting the index $x_{i}$. (b) is obtained from averaging over the discrete random variable $G_{\mathrm{T}}$ that corresponds to the directionality gain of the interfering link where $p_{k}$ and $G_{k}$ are given in Table~\ref{table:antenna} and from the moment generating functional (MGF) of the small scale fading gain $\chi_{\mathrm{L}}$ modeled as a gamma distribution. Finally, (c) follows from substituting $d_{\mathrm{L}}$ with $y$ and averaging over $y$ where $f_{Y_{\mathrm{L}}}(y,x_{\mathrm{L}})$ is the distance distribution from an interfering LOS THz AP located further that $x_{\mathrm{L}}$ and is given in~\cite[Lemma~$4$]{Wang} as $f_{Y_{\mathrm{L}}}(y,x)=\frac{f_{\mathrm{Z}}(y)\kappa_{\mathrm{L}}(y)}{\int_{x}^{z_{p}}f_{\mathrm{Z}}(z)\kappa_{\mathrm{L}}(z)\mathrm{d}z}$. Similarly, for a NLOS THz AP:
\begin{equation}
	\mathbb{E}_{P_{\mathrm{N},x_{i}}^{r}}\left[\exp\left(-s P_{\mathrm{N},x_{i}}^{r}\right)\right]=\sum_{k=1}^{4}p_{k}\int_{E_{\mathrm{LN}}(x_{\mathrm{L}})}^{z_{p}}\left(1+\frac{s P_{\mathrm{T}}\gamma_{\mathrm{T}}G_{k}e^{k_{a}(f_{\mathrm{T}})y}y^{-\alpha_{\mathrm{N}}}}{m_{\mathrm{N}}}\right)^{-m_{\mathrm{N}}}f_{Y_{\mathrm{N}}}(y,E_{\mathrm{LN}}(x_{\mathrm{L}}))\mathrm{d}y,
	\label{eq:EPNr_proof}
	\vspace{-0.3cm}
\end{equation}
where $f_{Y_{\mathrm{N}}}(y,E_{\mathrm{LN}}(x_{\mathrm{L}}))=\frac{f_{\mathrm{Z}}(y)\kappa_{\mathrm{N}}(y)}{\int_{E_{\mathrm{LN}}(x_{\mathrm{L}})}^{z_{p}}f_{\mathrm{Z}}(z)\kappa_{\mathrm{N}}(z)\mathrm{d}z}$ is the distance distribution from a NLOS THz interfering AP located further than $E_{\mathrm{LN}}(x_{\mathrm{L}})$. By plugging (\ref{eq:EPLr_proof}), (\ref{eq:EPNr_proof}) and (\ref{eq:EIL_proof}) in (\ref{eq:LIL_proof}), we can get the final expression in (\ref{eq:LIL}). 
\vspace{-0.5cm}
\subsection{Proof of Theorem~\ref{theorem:rate}}\label{app:rate}
Given the Shannon's bound for the instantaneous $\mathrm{SINR}$ with a transmission bandwidth $W$, the average achievable rate for the DL is given by
\begin{equation}\small
\begin{aligned}
    \tau&=\mathbb{E}\left[W\log_2(1+\mathrm{SINR})\right]\stackrel{(a)}{=}\int_{0}^{\infty}\mathbb{P}\left[W\log_2(1+\mathrm{SINR})>y\right]\mathrm{d}y
    \stackrel{(b)}{=}W_{\mathrm{T}}\int_{0}^{\infty}\mathbb{P}\left[\log_{2}(1+\mathrm{SINR})>y|C_{\mathrm{L}}\right]\mathrm{d}y A_{\mathrm{L}}\\&+W_{\mathrm{T}}\int_{0}^{\infty}\mathbb{P}\left[\log_{2}(1+\mathrm{SINR})>y|C_{\mathrm{N}}\right]\mathrm{d}y A_{\mathrm{N}} + W_{\mathrm{R}}\int_{0}^{\infty}\mathbb{P}\left[\log_{2}(1+\mathrm{SINR})>y|C_{\mathrm{R}}\right]\mathrm{d}y A_{\mathrm{R}},
\end{aligned}
\vspace{-0.3cm}
\end{equation}
where $W_{\mathrm{T}}$ and $W_{\mathrm{R}}$ are the transmission bandwidths for THz and RF. (a) follows from $\mathbb{E}\left[ X\right]=\int_{0}^{\infty}\mathbb{P}\left[X>y\right]\mathrm{d}y$, and
(b) is obtained by referring to the law of total probability and the linearity of integrals. $A_{\mathrm{L}}$, $A_{\mathrm{N}}$ and $A_{\mathrm{R}}$ denote the association probabilities given in (\ref{eq:AL}), (\ref{eq:AN}) and (\ref{eq:AR}), respectively. Now, given that the reference UE is associated with a LOS THz AP, the conditional average rate $\tau_{\mathrm{L}}$ is given by
\begin{equation}\small
\begin{aligned}
   \tau_{\mathrm{L}}&=W_{\mathrm{T}}\int_{0}^{\infty}\mathbb{P}\left[\log_{2}\left(1+\mathrm{SINR}_{\mathrm{L}}\right)>y\right]\mathrm{d}y=\frac{W_{\mathrm{T}}}{\ln{2}} \int_{0}^{\infty}\mathbb{P}\left[\mathrm{SINR}_{\mathrm{L}}>e^{y}-1\right]\mathrm{d}y\\
   &\stackrel{(a)}{=}\frac{W_{\mathrm{T}}}{\ln{2}}\int_{0}^{\infty}\frac{1}{1+t}\mathbb{P}\left[\frac{P_{\mathrm{T}} \gamma_{\mathrm{T}} G_{\mathrm{T},0} e^{-k_a(f_{\mathrm{T}})x_{\mathrm{L}}}x_{\mathrm{L}}^{-\alpha_{\mathrm{L}}}\chi_{\mathrm{L},0}}{I_{\mathrm{L}}+\sigma_{\mathrm{T}}^2}>t\right]\mathrm{d}t\\
   &\stackrel{(b)}{=}\frac{W_{\mathrm{T}}}{\ln{2}}\int_{0}^{\infty}\frac{1}{1+t}\mathbb{E}_{G_{\mathrm{T},0},x_{\mathrm{L}},I_{\mathrm{L}}}\left[\mathbb{P}\left[\chi_{\mathrm{L},0}>\frac{t\left(I_{\mathrm{L}}+\sigma_{\mathrm{T}}^2\right)}{P_{\mathrm{T}}\gamma_{\mathrm{T}}G_{\mathrm{T},0}e^{-k_{a}(f_{\mathrm{T}})x_{\mathrm{L}}}x_{\mathrm{L}}^{-\alpha_{\mathrm{L}}}}\right]\right]\mathrm{d}t,
\end{aligned}
\end{equation}
where (a) follows from the change of variable $t=e^y-1$ and from the expression of $\mathrm{SINR}_{\mathrm{L}}$ given in (\ref{eq:SINR_L}) and (b) follows from taking the expectation over $x_{\mathrm{L}}$, $I_{\mathrm{L}}$ and $G_{\mathrm{T},0}$. The proof proceeds following the same steps of Theorem~\ref{theorem:coverage}, therefore we omit it here. Finally, the conditional average achievable rates given that the reference UE is associated with a NLOS THz AP and with an RF AP can also be derived by following the same proof as that of LOS THz AP association.
\vspace{-0.5cm}
\bibliography{IEEEabrv,references}

\begin{thebibliography}{10}

\bibitem{Andrews}
J.~G. Andrews, S.~Buzzi, W.~Choi, S.~V. Hanly, A.~Lozano, A.~C.~K. Soong, and
  J.~C. Zhang, ``What will {5G} be?,'' {\em IEEE J.~Sel.~Areas in Commun.},
  vol.~32, no.~6, pp.~1065--1082, 2014.

\bibitem{Sarieddeen}
H.~Sarieddeen, N.~Saeed, T.~Y. Al-Naffouri, and M.-S. Alouini, ``Next
  generation terahertz communications: A rendezvous of sensing, imaging, and
  localization,'' {\em IEEE Commun. Mag.}, vol.~58, no.~5, pp.~69--75, 2020.

\bibitem{Elayan}
H.~Elayan, O.~Amin, B.~Shihada, R.~M. Shubair, and M.-S. Alouini, ``Terahertz
  band: The last piece of {RF} spectrum puzzle for communication systems,''
  {\em IEEE Open Journal of the Communications Society}, vol.~1, pp.~1--32,
  2020.

\bibitem{rajatheva2020scoring}
N.~Rajatheva, I.~Atzeni, S.~Bicais, E.~Bjornson, A.~Bourdoux, S.~Buzzi,
  C.~D'Andrea, J.-B. Dore, S.~Erkucuk, M.~Fuentes, {\em et~al.}, ``Scoring the
  terabit/s goal: {B}roadband connectivity in {6G},'' {\em arXiv preprint
  arXiv:2008.07220}, 2020.

\bibitem{Zhengquan}
Z.~Zhang, Y.~Xiao, Z.~Ma, M.~Xiao, Z.~Ding, X.~Lei, G.~K. Karagiannidis, and
  P.~Fan, ``{6G} wireless networks: Vision, requirements, architecture, and key
  technologies,'' {\em {IEEE} Veh. Technol. Mag.}, vol.~14, no.~3, pp.~28--41,
  2019.

\bibitem{Lima9330512}
C.~De~Lima, D.~Belot, R.~Berkvens, A.~Bourdoux, D.~Dardari, M.~Guillaud,
  M.~Isomursu, E.-S. Lohan, Y.~Miao, A.~N. Barreto, M.~R.~K. Aziz,
  J.~Saloranta, T.~Sanguanpuak, H.~Sarieddeen, G.~Seco-Granados, J.~Suutala,
  T.~Svensson, M.~Valkama, B.~Van~Liempd, and H.~Wymeersch, ``Convergent
  communication, sensing and localization in {6G} systems: {An} overview of
  technologies, opportunities and challenges,'' {\em IEEE Access}, vol.~9,
  pp.~26902--26925, 2021.

\bibitem{Ian}
I.~F. Akyildiz, J.~M. Jornet, and C.~Han, ``Terahertz band: Next frontier for
  wireless communications,'' {\em Physical Communication}, vol.~12, pp.~16--32,
  2014.

\bibitem{sarieddeen2021overview}
H.~Sarieddeen, M.-S. Alouini, and T.~Y. Al-Naffouri, ``An overview of signal
  processing techniques for terahertz communications,'' {\em Proceedings of the
  IEEE}, 2021.

\bibitem{tarboush2021teramimo}
S.~Tarboush, H.~Sarieddeen, H.~Chen, M.~H. Loukil, H.~Jemaa, M.~S. Alouini, and
  T.~Y. Al-Naffouri, ``{TeraMIMO}: A channel simulator for wideband
  ultra-massive {MIMO} terahertz communications,'' {\em arXiv preprint
  arXiv:2104.11054}, 2021.

\bibitem{Jornet}
J.~M. Jornet and I.~F. Akyildiz, ``Channel modeling and capacity analysis for
  electromagnetic wireless nanonetworks in the terahertz band,'' {\em {IEEE}
  Trans. Wireless Commun.}, vol.~10, no.~10, pp.~3211--3221, 2011.

\bibitem{Han}
C.~Han, A.~O. Bicen, and I.~F. Akyildiz, ``Multi-wideband waveform design for
  distance-adaptive wireless communications in the terahertz band,'' {\em IEEE
  Trans.~Signal Processing}, vol.~64, no.~4, pp.~910--922, 2016.

\bibitem{Akyildiz}
I.~F. Akyildiz, C.~Han, and S.~Nie, ``Combating the distance problem in the
  millimeter wave and terahertz frequency bands,'' {\em IEEE Commun. Mag.},
  vol.~56, no.~6, pp.~102--108, 2018.

\bibitem{Chen}
Y.~Chen, Y.~Li, C.~Han, Z.~Yu, and G.~Wang, ``Channel measurement and
  ray-tracing-statistical hybrid modeling for low-terahertz indoor
  communications,'' {\em {IEEE} Trans. Wireless Commun.}, 2021.

\bibitem{Faisal}
A.~Faisal, H.~Sarieddeen, H.~Dahrouj, T.~Y. Al-Naffouri, and M.-S. Alouini,
  ``Ultramassive {MIMO} systems at terahertz bands: Prospects and challenges,''
  {\em {IEEE} Veh. Technol. Mag.}, vol.~15, no.~4, pp.~33--42, 2020.

\bibitem{ElSawy}
H.~ElSawy, A.~Sultan-Salem, M.-S. Alouini, and M.~Z. Win, ``Modeling and
  analysis of cellular networks using stochastic geometry: A tutorial,'' {\em
  {IEEE} Commun. Surveys Tuts.}, vol.~19, no.~1, pp.~167--203, 2017.

\bibitem{Kokkoniemi}
J.~{Kokkoniemi}, J.~{Lehtomäki}, and M.~{Juntti}, ``Stochastic geometry
  analysis for mean interference power and outage probability in {THz}
  networks,'' {\em {IEEE} Trans. Wireless Commun.}, vol.~16, no.~5,
  pp.~3017--3028, 2017.

\bibitem{Petrov}
V.~Petrov, M.~Komarov, D.~Moltchanov, J.~M. Jornet, and Y.~Koucheryavy,
  ``Interference and {SINR} in millimeter wave and terahertz communication
  systems with blocking and directional antennas,'' {\em {IEEE} Trans. Wireless
  Commun.}, vol.~16, no.~3, pp.~1791--1808, 2017.

\bibitem{Yao2}
X.-W. Yao, C.-C. Wang, W.-L. Wang, and C.~Han, ``Stochastic geometry analysis
  of interference and coverage in terahertz networks,'' {\em Nano Communication
  Networks}, vol.~13, pp.~9--19, 2017.

\bibitem{Wang}
C.-C. Wang, X.-W. Yao, C.~Han, and W.-L. Wang, ``Interference and coverage
  analysis for terahertz band communication in nanonetworks,'' in {\em
  Proc.~IEEE Global Telecommun.~Conf. (GLOBECOM)}, pp.~1--6, 2017.

\bibitem{Dmitri}
D.~Moltchanov, P.~Kustarev, and Y.~Koucheryavy, ``Analytical approximations for
  interference and sir densities in terahertz systems with atmospheric
  absorption, directional antennas and blocking,'' {\em Physical
  Communication}, vol.~26, pp.~21--30, 2018.

\bibitem{Humadi}
K.~Humadi, I.~Trigui, W.-P. Zhu, and W.~Ajib, ``Coverage analysis of
  user-centric dense terahertz networks,'' {\em {IEEE} Commun. Lett.}, 2021.

\bibitem{Huq}
K.~M.~S. Huq, J.~Rodriguez, and I.~E. Otung, ``{3D} network modeling for
  {THz}-enabled ultra-fast dense networks: A {6G} perspective,'' {\em IEEE
  Commun.~Standards Mag.}, vol.~5, no.~2, pp.~84--90, 2021.

\bibitem{Wu}
Y.~{Wu} and C.~{Han}, ``Interference and coverage analysis for indoor terahertz
  wireless local area networks,'' in {\em IEEE Globecom Workshops (GC Wkshps)},
  pp.~1--6, 2019.

\bibitem{Wu2}
Y.~Wu, J.~Kokkoniemi, C.~Han, and M.~Juntti, ``Interference and coverage
  analysis for terahertz networks with indoor blockage effects and
  line-of-sight access point association,'' {\em {IEEE} Trans. Wireless
  Commun.}, vol.~20, no.~3, pp.~1472--1486, 2021.

\bibitem{shafie}
A.~Shafie, N.~Yang, S.~Durrani, X.~Zhou, C.~Han, and M.~Juntti, ``Coverage
  analysis for {3D} terahertz communication systems,'' {\em IEEE J.~Sel.~Areas
  in Commun.}, vol.~39, no.~6, pp.~1817--1832, 2021.

\bibitem{Shafie2}
A.~Shafie, N.~Yang, Z.~Sun, and S.~Durrani, ``Coverage analysis for {3D}
  terahertz communication systems with blockage and directional antennas,'' in
  {\em Proc.~IEEE Int.~Conf.~Commun. Workshops (ICC Workshops)}, pp.~1--7,
  2020.

\bibitem{Sayehvand}
J.~{Sayehvand} and H.~{Tabassum}, ``Interference and coverage analysis in
  coexisting {RF} and dense terahertz wireless networks,'' {\em {IEEE} Wireless
  Commun. Lett.}, vol.~9, no.~10, pp.~1738--1742, 2020.

\bibitem{Shi}
M.~Shi, X.~Gao, A.~Meng, and D.~Niyato, ``Coverage and area spectral efficiency
  analysis of dense terahertz networks in finite region,'' {\em China
  Communications}, vol.~18, no.~5, pp.~120--130, 2021.

\bibitem{Raja}
A.~A. Raja, H.~Pervaiz, S.~A. Hassan, S.~Garg, M.~S. Hossain, and M.~J. Piran,
  ``Coverage analysis of {mmWave} and {THz}-enabled aerial and terrestrial
  heterogeneous networks,'' {\em {IEEE} Trans. Intell. Transp. Syst.},
  pp.~1--14, 2021.

\bibitem{3gpp38900}
3GPP, ``{Study on channel model for frequency spectrum above 6 {GHz} (Release
  14)},'' 3GPP TR 38.900 V14.2.0, June 2017.

\bibitem{sarieddeen2019terahertz}
H.~Sarieddeen, M.-S. Alouini, and T.~Y. Al-Naffouri, ``Terahertz-band
  ultra-massive spatial modulation {MIMO},'' {\em {IEEE} J. Sel. Areas
  Commun.}, vol.~37, no.~9, pp.~2040--2052, 2019.

\bibitem{Wildman}
J.~{Wildman} {\em et~al.}, ``On the joint impact of beamwidth and orientation
  error on throughput in directional wireless poisson networks,'' {\em {IEEE}
  Trans. Wireless Commun.}, vol.~13, no.~12, pp.~7072--7085, 2014.

\bibitem{Chetlur}
V.~V. {Chetlur} and H.~S. {Dhillon}, ``Downlink coverage analysis for a finite
  {3-D} wireless network of unmanned aerial vehicles,'' {\em IEEE
  Trans.~Commun.}, vol.~65, no.~10, pp.~4543--4558, 2017.

\end{thebibliography}
\bibliographystyle{ieeetr}
\end{document}